%% file: main_TSP.tex
\documentclass[journal]{IEEEtran} 

\usepackage{pgfplots}
\usepackage{pgfplotstable}
\usepackage{tikz}
\usepackage{caption} 
\usepackage{multirow}
\usepackage{colortbl} 
\usepackage{xcolor}

\pgfplotsset{compat=newest}

\input{Preamble.tex}

\usepackage[all=normal,paragraphs=tight,floats=normal,mathspacing=normal,wordspacing=tight,charwidths=tight,mathdisplays=normal,leading=normal]{savetrees}

\begin{document}

\title{Memory-Efficient Distributed Unlearning}

\author{Natalie Lang$^*$, Alon Helvits$^*$, and
Nir Shlezinger~\thanks{$^*$Equal contribution. The authors are with the School of ECE, 
Ben-Gurion University of the Negev
Be’er-Sheva, Israel. Emails: 
\texttt{\{langn,alonhel\}@post.bgu.ac.il, nirshl@bgu.ac.il}}}



\maketitle

\begin{abstract}
Machine unlearning considers the removal of the contribution of a set of data points from a trained model. In a distributed setting, where a server orchestrates training using data available at a set of remote users, unlearning is essential to cope with late-detected malicious or corrupted users. Existing distributed unlearning algorithms require the server to store all model updates observed in training, leading to immense storage overhead for preserving the ability to unlearn. 
In this work we study lossy compression schemes for facilitating distributed server-side unlearning with limited memory footprint. We propose {\em \ac{medu}}, a  hierarchical lossy compression scheme tailored for server-side unlearning, that integrates user sparsification, differential thresholding, and  random lattice coding, to substantially reduce memory footprint. We rigorously analyze \ac{medu}, deriving an upper bound on the difference between the desired model that is trained from scratch and the model unlearned from lossy compressed stored updates. Our bound outperforms the state-of-the-art known bounds for non-compressed decentralized server-side unlearning, even when lossy compression is incorporated. We further provide a numerical study, which shows that suited lossy compression can enable distributed unlearning with notably reduced memory footprint at the server while preserving the utility of the unlearned model.
\end{abstract}

\begin{IEEEkeywords}
Machine unlearning, federated learning, lossy compression, quantization, user-selection.
\end{IEEEkeywords}

\acresetall

\section{Introduction}\label{sec:intro}
\IEEEPARstart{D}{eep} learning  usually requires large volumes of training data to result in high-performance models. While data is often abundantly available in the `big data' era~\cite{jordan2015machine}, the source of the data might raise privacy or ownership concerns~\cite{liu2021machine}, among which is the GDPR {\em \ac{rtbf}} \cite{voigt2017eu}, as well as security concerns, as adversaries can maliciously modify the training (poisoning or backdoor attacks) or test data (evasion attacks)~\cite{pitropakis2019taxonomy}.
To address those, the paradigm of {\em machine unlearning} aims to selectively remove the influence of certain data points from a trained model, with neither retraining it from scratch nor impacting its original performance and predictive power \cite{bourtoule2021machine,shaik2024exploring}. 

In distributed learning systems, such as \ac{fl}~\cite{mcmahan2017communication,kairouz2021advances, gafni2022federated}, where training is done on edge devices without data sharing, the ability to unlearn is often essential. An important use case is coping with  late detection of a malicious user, possibly after learning is concluded~\cite{wang2023federated, nguyen2022survey}. 
Distributed unlearning frameworks are categorized based on the identity of the users participating in the unlearning procedure~\cite{liu2023survey}. {\em Server-side unlearning}, also termed {\em passive unlearning}, is the challenging setting in which only the server, who originally orchestrated the distributed learning procedure, participates in unlearning, as in the late detection use case~\cite{romandini2024federated}. 

Distributed unlearning brings forth new challenges compared to conventional centralized setups \cite{fraboni2024sifu, huynh2024fast, tao2024communication,cao2023fedrecover}. This is because the user to be forgotten may be a late-discovered  Byzantine adversary that cannot be trusted for further collaboration; Alternatively, the server can be left with no connected devices, for which  training the model anew with all the rest of users (the {\em train-from-scratch} model) is out of reach~\cite{liu2023survey}. 

Various distributed unlearning algorithms have been proposed to tackle these potential scenarios.
In FedRecovery \cite{zhang2023fedrecovery},  in addition to retaining the clients historical data, the server also quantifies their contributions based on gradient residuals. Upon an unlearning request, the server removes the unlearned user past
contributions through a fine-tuning process. A more efficient version is then suggested by
Crab \cite{jiang2024towards}, which uses only selective historical information and further assists a less-degraded historical model than the initial one. 
The recovery process can be improved by introducing constraints, e.g., a penalty term based on projected gradients \cite{fu2024client, shao2024federated}; randomly initialized degradation models \cite{zhao2023federated}; estimated skew \cite{huynh2024fast}; 
and retraining based on the change of sampling probability \cite{tao2024communication}. In VeriFi \cite{gao2024verifi}, the target client collaborates with the server and marks his data to verify the unlearning. 
As means to preserve performance despite the target client contribution elimination, knowledge distillation was shown to facilitate information transfer from the trained model to the unlearned one \cite{wu2022federated, wu2023unlearning}.

It is noted that the surveyed works are either not suitable to server-side unlearning or heavily relying on discarding historical parameter updates of the removed user. Accordingly, the server that orchestrates the learning procedure has to store all past contributions of all clients in order to have the ability to unlearn when required~\cite{gao2024verifi}. This induces a notable limitation, necessitating excessive and possibly prohibitive overhead for saving a large number of highly parameterized updates for each user in each global training round. It is concluded that distributed unlearning is challenging in general, and once  constrained to server-side unlearning, its associated state-of-the-art proposed methods result in substantial memory footprint. This, in turn, motivates the formulation of dedicated algorithms which aim to address the storage challenge in server-side unlearning.

From a signal processing perspective, memory footprint can be reduced by saving (lossy) compressed versions of the quantities rather than high-precision ones.
While lossy compression techniques are still unexplored for distributed unlearning, various schemes have been considered for distributed training, particularly in the aim of alleviating uplink communication bottlenecks~\cite{chen2020joint, li2020federated,chen2021communication}. Among the existing methods considered for communication efficient distributed learning are sub-sampling or sparsification \cite{lin2017deep, hardy2017distributed, aji2017sparse,konevcny2016federated,stich2018sparsified,alistarh2018convergence,han2020adaptive}; and probabilistic scalar \cite{wen2017terngrad, alistarh2017qsgd, horvoth2022natural, reisizadeh2020fedpaq, horvath2023stochastic} or vector quantization \cite{lang2023joint, azimi2024federated}, as well as voting-type learning~\cite{lang2023cpa}. 
Lossy compression, as opposed to its lossless counterpart, inevitably induces distortion, yet enables substantial memory savings~\cite{polyanskiy2014lecture}. Nevertheless, the random  distortion induced by probabilistic lossy compression can be rendered to have a negligible effect on the learning procedure~\cite{alistarh2017qsgd, shlezinger2020uveqfed}.  

Inspired by that, the goal of this work is to alleviate the storage burden introduced in server-side unlearning using {\em lossy source coding} tools. 
To do so, we propose {\em \ac{medu}}, in which  the server does not save the users' sent  model updates, but rather their lossy compressed versions, considerably reducing its memory footprint. We  introduce a hierarchical lossy compression scheme tailored for server-side unlearning, based on user sparsification,  differential thresholding, and (probabilistic) lattice quantization. The reduction in storage and effect of the induced distortion of \ac{medu} on the unlearned model is further analytically and experimentally analyzed using conventional metrics in the unlearning literature. It is then revealed that while integrating proper compression into unlearning significantly relieves the server's storage load, it does not change the asymptotic behavior of the unlearned model, while effectively removing the influence of the unlearned user. 

Our main contributions are summarized as follows:
\begin{itemize}
    \item {\bf Compressed Unlearning Framework:} We study lossy compression in a distributed (federated) unlearning framework, and identify the main considerations to mitigate the memory footprint accumulated over the learning procedure. We are, to the best of our knowledge, the first work that systematically examines the theoretical and numerical aspects of compressed distributed machine unlearning. 
    \item {\bf Hierarchical Compression Scheme:} We introduce \ac{medu}, a hierarchical lossy compression method that notably reduces the memory footprint of distributed unlearning and balances the quality of unlearning with storage limitations. 
    \item {\bf Theoretical Analysis:} We theoretically analyze  gradient descent based distributed learning using \ac{medu}. We derive an upper bound on the commonly adopted proximity to the desired train-from-scratch model of server-side unlearning.
    \item {\bf Asymptotic Convergence}: We show that in the asymptotic regime, for growing amount of gradient descent iterations, the compressed unlearned model can be made not to diverge. Our characterization specializes also non-compressed distributed unlearning, and improves the state-of-the-art asymptotic behavior of the bound presented in~\cite{huynh2024fast}. 
    \item {\bf Extensive Experimentation}: \ac{medu} and its integrated compression mechanisms are quantitatively validated in a \ac{fu} setup,
    evaluated using an established backdoor attack. We demonstrate that \ac{medu} notably reduces storage overhead in unlearning, with only a minor degradation in accuracy and while preserving the ability to defend against late discovered adversaries. 
\end{itemize}

The rest of this paper is organized as follows: Section~\ref{sec:sys_model_prelim} briefly reviews  both \ac{fl} and \ac{fu}, as well as related preliminaries in lossy compression. Section~\ref{sec:method} introduces our compressed distributed unlearning framework, while Section~\ref{subsec:analysis} theoretically analyzes its properties and guarantees, followed by the derivation of a convergence bounds. Our numerical study is given in Section~\ref{sec:experiments}, while Section~\ref{sec:conclusions} provides concluding remarks.

Throughout this paper, we use boldface lower-case letters for vectors, e.g., $\vx$. We use calligraphic letters for sets, e.g., $\cX$, with $|\cX|$ being its cardinality. The stochastic expectation, variance, inner product, and $\ell_2$ norm are denoted by $\E[\cdot]$, $\Var(\cdot)$, $\langle\cdot;\cdot\rangle$, and $\|\cdot\|$, respectively; while $\sR$ is the set of real numbers. 

\section{System Model}\label{sec:sys_model_prelim}
Here, we formulate compressed unlearning,  reviewing \ac{fl} and \ac{fu} in Subsections~\ref{subsec:FL}-\ref{subsec:FU} respectively. 
For clarity, we summarize the symbols used in this section in Table~\ref{tbl:notations}. 

\begin{table}
\caption{Summary of notations.}
\label{tbl:notations}
\centering
\begin{adjustbox}{width=\columnwidth} 
\begin{tabular}{|c|c|c|}
\hline
{\bf Symbol} & {\bf Description} & {\bf Definition} \\
\hline
$u, U$ & index and number of edge clients &\\
$t, T$ & index and number of global learning rounds&\\
$M$ & number of model parameters &\\
$\eta_t$ & learning rate at time $t$ &\\
$\vw^\star$ & \ac{fl} optimal global model & \eqref{eq:FL_optimal_model}\\
$\vg_t^u$ & stochastic gradient of user $u$ at round $t$ & \eqref{eq:lsgd}\\
$\vw_t$ & \ac{fl} global model at time $t$ & \eqref{eq:FedAvg_update}\\
$\vw^\star_t$ & desired train-from-scratch unlearned model & \eqref{eq:train_from_scratch}\\
$\vw'_t$ & generic server-side unlearning & \eqref{eq:unlearned_model}\\
\hline
\end{tabular}
\end{adjustbox}
\end{table} 

\subsection{Distributed Learning}\label{subsec:FL}
We consider a learning setup where a server trains a model with parameters $\vw\in \sR^M$ using data available at $U$ users,  indexed by $ u \in [U] :=\{1,\dots, U\}$. Let $\cL_u(\vw)$ denote the $u$th user empirical risk, which is computed over its local dataset $\cD_u$, the desired model is the minimizer of $\{\cL_u(\vw)\}$ average, that is 
\begin{align}\label{eq:FL_optimal_model}
    \vw^\star:=\argmin_{\vw}\Big\{\cL(\vw):=\frac{1}{U}\sum_{u=1}^U \cL_u(\vw)\Big\}.
\end{align}
Unlike conventional centralized learning, the datasets $\{\cD_u\}$ are not shared with the server due to, e.g., privacy considerations, and thus the learning is federated~\cite{kairouz2021advances}, and operates in multiple rounds.

For every round $t$, the server distributes the global model $\vw_t$ (where $\vw_0$ is the vector of initial weights) to the users who set $\{\vw^u_t=\vw_t\}_u$. Each user locally performs a training iteration using its local $\cD_u$, to update $\vw^u_t$ into $\vw^u_{t+1}$ via the highly adopted \ac{sgd} \cite{stich2018local}:
\begin{align}\label{eq:lsgd}
   \vw^u_{t+1}:= \vw_t - \eta_t\vg_t^u, \quad \vg_t^u:=\nabla\cL_u(\vw_t; \xi^u_t);
\end{align}
where $\eta_t$ is the learning rate (step size) and $\xi^u_t$ is a sample uniformly chosen from the local data. 

Then, each edge device shares the model update (gradient) $\nabla\cL_u(\vw_t; \xi^u_t)$ with the server, which in turn aggregates all to form an updated global model via the arguably most common \ac{fl} scheme of~\cite{mcmahan2017communication} -- \ac{fa}:
\begin{align}\label{eq:FedAvg_update}
    \vw_{t+1} := \frac{1}{U}\sum_{u=1}^U \vw^u_{t+1}=\vw_t - \eta_t\frac{1}{U}\sum_{u=1}^U \vg_t^u.
\end{align}
For simplicity, \eqref{eq:FedAvg_update} is formulated with all users participating in each round, which  straightforwardly extends to partial user participation~\cite{li2019convergence} and selection~\cite{peleg2025pause}. The updated global model is again distributed to the users, and the learning procedure continues. The above steps are summarized as Algorithm~\ref{alg:fl}.

\SetKwBlock{Users}{users side:}{end}
\SetKwBlock{Server}{server side:}{end}
\SetKwBlock{DoParallel}{do in parallel for $u \in \{1,\dots,U\}$}{end}
\begin{algorithm}
\caption{\ac{fl} at round $t$}
\label{alg:fl}
\Users{
\DoParallel{
Set $\vw^u_t=\vw_t$ and update into $\vw^u_{t+1}$ via~\eqref{eq:lsgd};\\
Send to server the gradient $\vg_t^u$;\\
}}    
\Server{
Update $\vw_t$ via \eqref{eq:FedAvg_update};\\
Distribute $\vw_{t+1}$ to all local users;}
\Return{Updated global model, $\vw_{t+1}$;}
\end{algorithm}

\subsection{Distributed Unlearning}\label{subsec:FU}
{\em Distributed unlearning} extends the distributed learning framework to ensure the \ac{rtbf} of its  users upon request, as well as the ability to eliminate  maliciously injected backdoors once they are revealed. The goal of unlearning here is to erase the contributions of a user (or a group of users) while preserving the performance of the  model acquired using the remaining clients~\cite{romandini2024federated}.

\smallskip
\subsubsection{Unlearning Task}
To formulate the unlearning paradigm, consider a distributed (federated) learning procedure that iterated over $T>1$ rounds up to the arrival of the unlearning request regarding the $\UnlUser$th user, $ \UnlUser\in [U]$.  
The desired unlearned model, coined the {\em train-from-scratch} model, is the one obtained by naively retraining the global model  using all users except for the omitted user $\UnlUser$ \cite{liu2023survey}, i.e., by iterating from $\vw^\star_0= \vw_0$ via 
\begin{align}\label{eq:train_from_scratch}
     \vw^\star_{T+1} = \vw^\star_T - \eta_T\frac{1}{U-1}\sum_{\substack{u=1\\ u \neq \UnlUser}}^U
     \nabla\cL_u(\vw^\star_T, \xi^u_t).
\end{align}

As elaborated above, retraining from scratch is often infeasible. Most existing \ac{fu} works, e.g., \cite{liu2021federaser,cao2023fedrecover,fraboni2024sifu}, relax it by balancing between partially retraining the local models and the subtraction of the unlearned user past updates, as both $\vw_{t+1},\vw^\star_{t+1}$ rely on accumulating model updates, according to \eqref{eq:FedAvg_update}, \eqref{eq:train_from_scratch}; respectively. 
Particularly, iterating over $T$ rounds of \eqref{eq:FedAvg_update} results in
\begin{align}\label{eq:iterating_FedAvg}
    \vw_{T+1} &= \vw_T - \eta_T\frac{1}{U}\sum_{u=1}^U \vg_T^u
            &= \vw_0 - \sum_{t=0}^T\eta_t\frac{1}{U}\sum_{u=1}^U \vg_t^u.
\end{align}
\subsubsection{Server-Side Unlearning}
{\em Server-side unlearning} refers to distributed unlearning that is carried out solely on the server-side, without users' retraining. In light of \eqref{eq:iterating_FedAvg}, when focusing on server-side unlearning, a generic unlearning rule is based on subtracting the unlearned user past gradients; as proposed in~\cite[Algorithms 1-2]{huynh2024fast}. There, when taking the pre-determined {\em skewness} parameter $\alpha=0$, the unlearned model is given by:   
\begin{align}\label{eq:unlearned_model}
\vw'_{T+1} &=  \vw_0 - \sum_{t=0}^{T} \eta_t\frac{1}{U-1} \sum_{\substack{u=1\\ u \neq \UnlUser}}^U \vg_t^u.
\end{align} 
We note that while \eqref{eq:unlearned_model} is designed assuming that the local objective functions are uniformly continuous and strongly convex, it was shown to also lead to reliable server-side unlearning in settings where these assumptions do not hold~\cite{huynh2024fast}.
The overall procedure, carrying out $T$ training rounds followed by server-side unlearning of user $\UnlUser$, is outlined as Algorithm~\ref{alg:fu}.

\begin{algorithm}
\caption{\ac{fl} + unlearning at round $T$}
\label{alg:fu}
\SetKw{Initialization}{initialization:}
\SetKwBlock{Server}{server side:}{end}
\SetKwBlock{DoParallel}{users $u \in [U]$ do in parallel}{end}
\Initialization{$\vw_0$, $T$;}\\
\For{$t\in\{0,\dots,T\}$}{
    Set $\vw_{t+1}$ via Algorithm~\ref{alg:fl};\\
\Server{
    Store $\{\vg_t^u\}_{u=1}^U$;\\
}}
\Server{
\If{unlearning for user $\UnlUser$}{
Compute $\vw'_{T+1}$ via \eqref{eq:unlearned_model};\\
\Return{Unlearned global model $\vw'_{T+1}$;}
}
}
\Return{Updated global model, $\vw_{T+1}$;}
\end{algorithm}
\smallskip
\subsubsection{Storage Requirements}\label{subsec:prebelm_def}
To be able to unlearn via \eqref{eq:unlearned_model}, the server must store all past local updates from all users during training~\cite{cao2023fedrecover}; i.e., storing $\{\vg_t^u\}$ for $1\leq u\leq U$ and $0\leq t\leq T$. When each of the $M$  parameters is represented using ${b}$ bits, \ac{fu} involves storing $U \cdot t$ sequences of $ M  \cdot b$ bits, i.e., the memory footprint is
\smallskip
\begin{equation}\label{eqn:StorageFU}
    \boxed{\text{Storage}_{\rm FU}= U\cdot (T+1) \cdot M \cdot {b} ~\text{[bits]}.}  
\end{equation}

\smallskip
The memory footprint formulation in \eqref{eqn:StorageFU} reveals the substantial burden associated with server-side unlearning. These memory requirements become prohibitive when training high-resolution  (large $b$) highly parameterized models (large $M$) over many rounds (large $T$)  with data from numerous users (large $U$). For instance, just to have the ability to unlearn MobileNetV2~\cite{sandler2018mobilenetv2}, considered to be relatively compact  ($M\approx 3.3\cdot 10^6$ parameters), at double precision ($b=64$ bits), after trained by a non-massive network of $U=16$ users over $T=300$ rounds, the server should store over a terabit of data. 

\section{Compressed Distributed Unlearning}\label{sec:method}
In this section we introduce a compressed distributed unlearning framework, that reduces the memory overhead at the server, while minimally affecting its ability to unlearn. 
We first formulate a generic setting for integrating lossy compression into distributed unlearning, and highlight its specific requirements in Subsection~\ref{subsec:Compression_framework}. 
Then, in Subsection~\ref{subsec:method}, we construct \ac{medu} designed to meet the identified demands. 

\subsection{Compressed Distributed Unlearning Framework}\label{subsec:Compression_framework}
\subsubsection{Lossy Compression Basics}
To mitigate the memory overhead, we introduce a  \ac{fu} framework,  allowing compact storage of $\{\vg_t^u\}$ by the server. To formulate this mathematically, we first recall the definition of lossy source codes~\cite{polyanskiy2014lecture}:
\begin{definition}[Lossy Source Code]\label{def:quantizer}
    A lossy source code with compression rate $R$, input size $M_x$, input alphabet $\cX^{M_x}$, and output  alphabet $\hat\cX^{M_x}$, consists of:
    $(i)$ An encoder which maps the input into a discrete index and $(ii)$ a decoder which maps each index into a codeword in $\hat\cX^{M_x}$; respectively given by
\begin{align*}
    e: \cX^{M_x} \mapsto \{0,\dots, 2^{M_x R}-1\}:= \cI,\quad d: \cI \mapsto \hat\cX^{M_x}.
\end{align*}
    An input $\vx \in \cX^{M_x}$ is thus mapped into $\hat\vx = d(e(\vx)) \in \hat\cX^{M_x}$.
\end{definition}
The performance of a lossy source code is characterized using its  rate ${R}$ and distortion, the latter commonly being the \ac{mse}, i.e., $\frac{1}{M_x}\E\big[\|\vx - \hat{\vx}\|^2\big]$. Such schemes can be combined with subsequent lossless compression, e.g., entropy coding~\cite{cover1999elements}, to further reduce storage requirements, without affecting the formulation of the recovered $\hat\vx$.

\begin{figure*}
\centering
\includegraphics[width=1.02\textwidth]{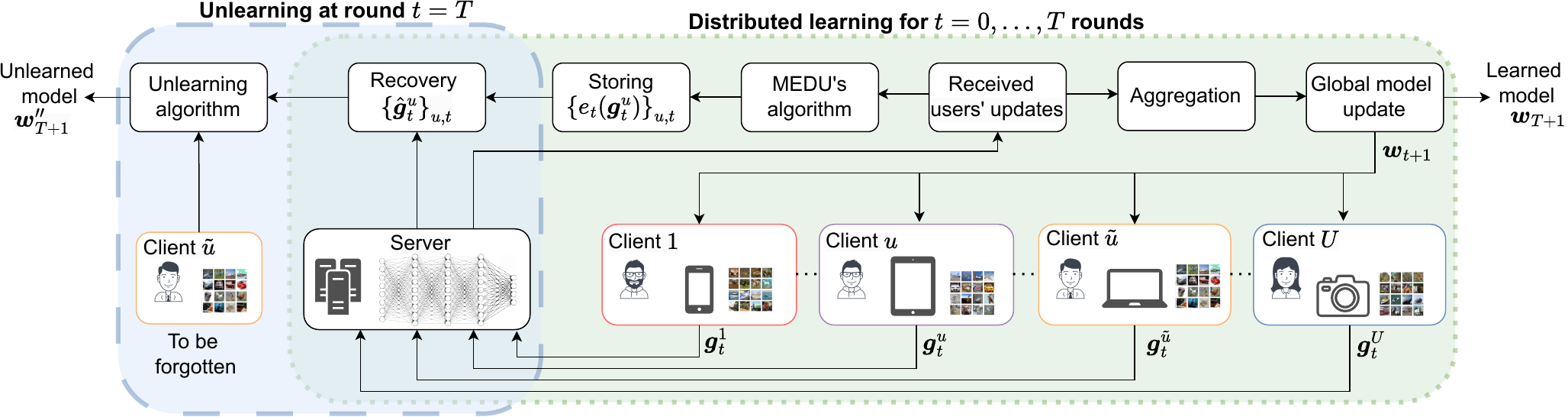}
\caption{\ac{medu}'s overview, describing \ac{fl} using $U$  clients, where the server stores compressed, with a subsequent unlearning of user $\UnlUser$.}
\label{fig:block_diagram}
\end{figure*}

\smallskip
\subsubsection{Problem Definition}
In the context of distributed unlearning, the data that is to be compressed is the model updates history. In this case, $\cX$ is the set of possible values of each weight (which we approximated as continuous, i.e., $\cX = \mathbb{R}$), while the number of weights that are to be compressed are $M_x=U\cdot T \cdot M$. 

Still, to facilitate implementation and scalability, we impose the following requirements on lossy source coding for \ac{fu}:
\begin{enumerate}[label={\em R\arabic*}]    
    \item \label{itm:realtime} The  encoder should operate separately on each round $t$.
        \item \label{itm:rate} The encoding function should be the same for all users.
            \item \label{itm:universal} The scheme  must be invariant to the distribution of $\vg_t^u$. 
\end{enumerate}
Requirement \ref{itm:realtime} is imposed since there is no prior knowledge after how many learning rounds would an unlearning request arrive, and to avoid having the server locally storing all past model updates to be jointly compressed after multiple rounds;
Requirement \ref{itm:rate} facilitates scalability of distributed learning, supporting numerous diverse users without altering the encoding procedure;
Requirement \ref{itm:universal} stems from the fact that the server is unlikely to have prior knowledge of the model parameters distribution, and thus we are interested in methods which are universal. 

Combining the above requirements, a compressed distributed unlearning setting employs model updates which are given by
    \begin{align}\label{eqn:CompModel}
        \{\hat\vg_t^u\} &= 
        d\left(e\left(\{\vg_t^u\}_{t,u=0,1}^{T,U}\right)\right)
        = d \left(\left\{e_t\left(\vg_t^u\right)\right\}_{t,u=0,1}^{T,U}\right).         
    \end{align}
The compressed model updates in \eqref{eqn:CompModel} should enable the server to approach the  train-from-scratch model \eqref{eq:train_from_scratch} via unlearning. Specifically, the unlearning rule  
 \eqref{eq:unlearned_model}  now becomes   
\begin{align}\label{eq:compressed_unlearned_model} 
\vw''_{T+1} &= 
\vw_0 - \sum_{t=0}^{T} \eta_t\frac{1}{U-1} \sum_{\substack{u=1\\ u \neq \UnlUser}}^U \hat{\vg}_t^u.
\end{align}
Accordingly, the unlearning Algorithm~\ref{alg:fu} is reformulated into the compressed Algorithm~\ref{alg:compressed fu}, visualized as Fig.~\ref{fig:block_diagram}.

\SetKw{Initialization}{initialization:}
\SetKwBlock{Server}{server side:}{end}
\SetKwBlock{DoParallel}{users $u \in \{1,\dots,U\}$ do in parallel}{end}
\begin{algorithm}
\caption{\ac{fl}~+~compressed unlearning at round $T$}
\label{alg:compressed fu}
\Initialization{$\vw_0$, $T$;}\\
\For{$t\in\{0,\dots,T\}$}{
Set $\vw_{t+1}$ via Algorithm~\ref{alg:fl};\\
\Server{
Store $\{e_t\left(\vg_t^u\right)\}_{u=1}^U$;\\
}}
\Server{
\If{unlearning for user $\UnlUser$}{
    Decode all encoded updates $\{\hat\vg_t^u\}_{t,u=0,1}^{T,U}$ via \eqref{eqn:CompModel};\\
    Compute $\vw''_{T+1}$ via \eqref{eq:compressed_unlearned_model};\\
\Return{Unlearned global model $\vw''_{T+1}$;}}}
\Return{Updated global model, $\vw_{T+1}$;}
\end{algorithm}

\subsection{MEDU}\label{subsec:method}
\begin{figure*}
\centering
\includegraphics[width=1.02\textwidth]{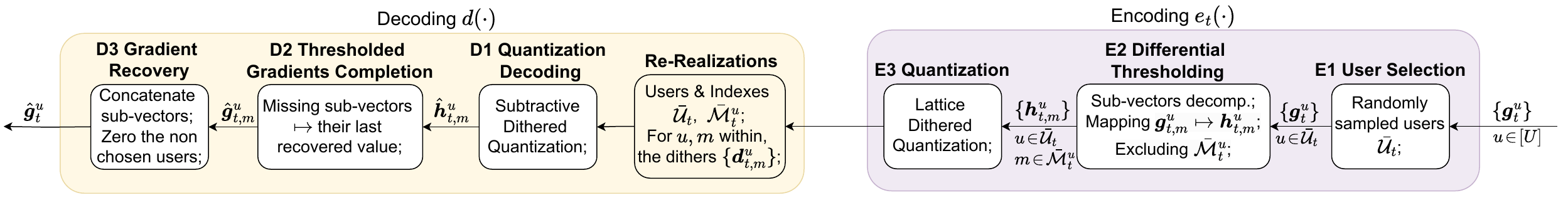}
\caption{The encoding the decoding schemes proposed by \ac{medu}, initialized during federated learning and unlearning, respectively.}
\label{fig:encoding-decosing_diagram}
\end{figure*}

Evidently, \ref{itm:realtime}-\ref{itm:universal} can be satisfied by any lossy source code that is invariant to the distribution of the model updates and operates separately at each round and for each user. However, to obtain a compression method that is simple to implement, while also being backed by theoretical guarantees, we opt a {\em hierarchical} scheme. Our proposed \ac{medu} scheme institutes a specific design for $e_t(\cdot)$ and $d(\cdot)$ in \eqref{eqn:CompModel} which gradually addresses each of the aspects determining the memory footprint of distributed unlearning in \eqref{eqn:StorageFU}; and is illustrated in Fig.~\ref{fig:encoding-decosing_diagram}. 

While most steps of the hierarchical compression require the realization of identical random quantities when encoding and decoding, we note that both procedures are carried out at the server side, though at different time steps, and thus one can reasonably assume shared source of common randomness. For instance, sharing of random quantities can be achieved via pseudo-random methods, that require storing a single seed used in random number generation~\cite{shlezinger2020uveqfed}, such that its excessive storage is made negligible. 

\smallskip
\subsubsection{Encoder}
The encoder operation gradually alleviates the memory overhead associated with the number of users ($U$), the number of rounds ($T$), the number of parameters ($M$), and the parameters' resolution ($b$). Its operation is divided into three steps:

\begin{enumerate}[label=\textbf{E\arabic*}, wide=0pt, leftmargin=*]  
\smallskip
\item {\bf User Selection.} \label{itm:E step User Selection}
Since the server has no prior on which user is to be unlearned, it has to store individual model updates for each of the users. However, the server can save only subsets of the users on each round, i.e., store $\bar{U} < U$ users. 
Based on the established guarantees of random  selection in \ac{fl}~\cite{li2019convergence}, we incorporate random user storage. 

The operation can be mathematically formulated as follows:
At the $t$th learning round, the server uniformly randomizes a $\bar{U}$ sized subset $\bar{\cU}_t\subset [U]$. Then, instead of storing all $U$ model updates, it stores $\{\vg_t^u\}_{u\in \bar{\cU}_t}$. Since the set $\bar{\cU}_t$ is randomized, there is no need to store it in encoding, as it can be regenerated upon decoding using, e.g., a stored seed.

\smallskip
\item {\bf Differential Thresholding.} \label{itm:E step Differential Thresholding}
The next encoding step accounts for the fact that for  convergent learning, the model updates are expected to exhibit decaying variations as learning progresses. Moreover, even prior to overall convergence, the variations in some of the parameters often saturate \cite{richter2021feature}. Therefore, for each of the users  in $\bar{\cU}_t$, the encoder only saves updates for  parameters which differ by more than $\delta_t$ from their previous stored value, where $\delta_t$ is a predefined threshold.

To formulate this precisely, let $\tau_u(t)$ be the last round in which the updates of user $u$ were stored, i.e., the maximal index not larger than $t$ for which $u \in \bar{\cU}_{\tau_u(t)}$.  Then, the  server divides each model update $\vg_t^u$ into a series of $M/L$ sub-vectors of length $L\times 1$, denoted $\{\vg_{t,m}^u\}_{m=1}^{M/L}$. The size of the sub-vectors $L$ is a hyperparameter related to the quantization mapping detailed in Step~\ref{itm:E step Quantization}. Each sub-vector is then mapped into a corresponding sub-vector $\vh_{t,m}^u$ via
\begin{equation}\label{eqn:DiffThresh}
    \vh_{t,m}^u = \begin{cases}
        \vc_{\rm last} & \left\|\vg_{t,m}^u - \vg_{{\tau_u(t)},m}^u \right\| < \delta_t, \\
   \vg_{t,m}^u & {\rm otherwise.}
    \end{cases}    
\end{equation}
In \eqref{eqn:DiffThresh}, $\vc_{\rm last}$ is a unique vector (e.g., whose entries all equal $-\infty$) indicating the server to not encode and store $\vg_{t,m}^u$ but rather use its last already-stored value in decoding, elaborated in Step~\ref{itm:D step grad completion}. We use $\bar{\cM}_t^u$ to denote the set of indices of $\vg_t^u$ sub-vectors that are directly stored:  
\begin{equation}\label{eq:indices directly stored}
    \bar{\cM}_t^u := \left\{m\in \{1,\ldots, M/L\} \ \middle| \ \vh_{t,m}^u \neq \vc_{\rm last}  \right\}.
\end{equation}
To inform the decoder which sub-vectors lie in $\bar{\cM}_t^u$, an additional string containing one bit per sub-vector is stored.

\smallskip
\item {\bf Quantization.} \label{itm:E step Quantization}
The final encoding step maps the sub-vectors 
$\big\{\{\vh_{t,m}^u\}_{u \in \bar{\cU}_t}\big\}_{m\in \bar{\cM}_t^u}$ into a bit stream stored by the server. 
Notably, the sub-vectors being directly stored in a digital form are only the ones in $\bar{\cM}_t^u$ for each chosen user in $\bar{\cU}_t$.

While quantizers can be optimized by tuning the code based on the input distribution, we are particularly interested in  universal quantizers, meeting \ref{itm:universal}.
A generic approach for universally choosing $L$-dimensional codewords is to realize them as the points of a {\em lattice}~\cite{zamir2014lattice}. We particularly adopt a probabilistic form of lattice quantization, known to yield noise-like distortion, termed {\em \ac{dq}}~\cite{gray1993dithered}:
\begin{definition}[Lattice \ac{dq}]
    Let $\mG$ be an $L\times L$ non-singular matrix and $\gamma$ is a truncated sphere radios. A lattice \ac{dq} maps  a dithered version of the input $\vx \in \sR^L$ into its nearest point in the lattice $\cP:=\{\mG\vz: \vz\in\sZ^L,\|\mG\vz\|<\gamma\}$, i.e.,
\begin{equation}\label{eq:lattice_quantizer}
    \argmin_{\vl\in\cP} \|\vx+\vd - \vl\|,   
\end{equation}
where $\vd$ is a zero-mean random dither signal.
\end{definition} 
When $\mG=\Delta \cdot\mI_{L\times L}$ for some $  \Delta>0$, \eqref{eq:lattice_quantizer} realizes  scalar uniform \ac{dq} per entry, known to be performance-wise inferior to its vectorized alternative\cite[Part V]{polyanskiy2014lecture}.
Proper selection of the distribution of $\vd$ can transform the quantization error into a form of noise that is uncorrelated with the input~\cite{gray1993dithered}. Following~\cite{zamir1996lattice}, we set $\vd$ to be uniformly distributed over the lattice basic cell (surrounding the zero point). Then, the sub-vectors $\{\vh_{t,m}^u\}_{m\in \bar{\cM}_t^u}$ are distorted by the random dither signals $\{\vd_{t,m}^u\}_{m\in \bar{\cM}_t^u}$ and quantized using \eqref{eq:lattice_quantizer}.
\end{enumerate}

\smallskip
\subsubsection{Decoder}
The decoder follows the hierarchical operation of the encoder, in an inverse manner. Accordingly, it operates sequentially from $t=1$ up to $t=T$. For each $t$, the first decoding step inverts the last encoding step, and so on, as elaborated next:

\begin{enumerate}[label=\textbf{D\arabic*}, wide=0pt, leftmargin=*]  
\smallskip
\item {\bf Quantization Decoding.} \label{itm:D Quantization Decoding}
The first decoding step recovers $\{\vh_{t,m}^u\}_{m\in \bar{\cM}_t^u}$. To that aim, the decoder first uses the $M/L$ bit string to obtain $\bar{\cM}_t^u$. The quantized vectors are then recovered via {\em \ac{sdq}}, namely, the server uses its seed to generate the dither signals $\{\vd_{t,m}^u\}_{m\in \bar{\cM}_t^u}$ and the set of stored users $\bar{\cU}_t$. The sub-vectors are then decoded  by subtracting the dither from their corresponding lattice point. As in \eqref{eqn:DiffThresh}, the remaining sub-vectors, i.e., with indices $u\in \bar{\cU}_t$ and $m\in \{1,\ldots M/L\}$, are decoded using the  codeword $\vc_{\rm last}$. Accordingly,
\begin{equation}\label{eqn:SDQ}
   \hspace{-0.7cm} \hat{\vh}_{t,m}^u \!=\!   
    \begin{cases}
    \mathop{\argmin}\limits_{\vl\in\cP} \|{\vh}_{t,m}^u \!+\! \vd_{t,m}^u\!- \!\vl\| \!-\! \vd_{t,m}^u & m \in \bar{\cM}_t^u,   \\     
    \vc_{\rm last} &  {\rm otherwise.} 
    \end{cases}    
\end{equation} 
 
\smallskip
\item {\bf Thresholded Gradients Completion.} \label{itm:D step grad completion}
Having recovered the set  $\{\vh_{t,m}^u\}$ for each user, the next steps replaces missing sub-vectors (marked with the codeword $\vc_{\rm last}$) with their last recovered value. Accordingly, for each user $u\in\bar{\cU}_t$, the server identifies the last update index $\tau_u(t)<t$, and sets the recovered gradient sub-vectors to be
\begin{equation}\label{eqn:DiffDec}
    \hat{\vg}_{t,m}^u =   
    \begin{cases}
    \hat{\vh}_{t,m}^u & m \in \bar{\cM}_t^u,\\        
    \hat{\vg}_{\tau_u(t),m}^u &   {\rm otherwise.}
    \end{cases}.    
\end{equation}

\smallskip
\item {\bf Gradient Recovery.}  \label{itm:D Gradient Recovery}
The final decoding step recovers the full gradients $ \hat{\vg}_{t}^u$  used for distributed unlearning via \eqref{eq:compressed_unlearned_model}. The recovered $M/L$ vectors $\{\hat{\vg}_{t,m}^u\}_m$ are concatenated to form the  $M\times 1$ gradient vector $ \hat{\vg}_{t}^u$ for each $u \in \bar{\cU}_t$. Since only $\bar{U} < U$ gradients are stored in each round, the full gradients are recovered via 
\begin{equation}\label{eqn:GradRec}
    \hat{\vg}_{t}^u
    = \begin{cases}
        \frac{U-1}{|\bar{\cU}_t/\{\UnlUser\}|} \left[\hat{\vg}_{t,1}^u,\ldots,\hat{\vg}_{t,M/L}^u \right] & u \in \bar{\cU}_t, \\
        \left[0,\ldots,0\right] &  {\rm otherwise.}
    \end{cases}    
\end{equation}
Note that during decoding, the server already knows the identity of the unlearned user, i.e., $\UnlUser$, and can thus compute \eqref{eqn:GradRec}.  
Under this setting, once all gradients up to round $T$ are recovered, the unlearning equation becomes 
\begin{align}\label{eq:compressed_unlearned_model2} 
\vw''_{T+1} &= 
\vw_0 - \sum_{t=0}^{T} \eta_t\frac{1}{|\bar{\cU}_t / \{\UnlUser\}| } \sum_{u \in \bar{\cU}_t / \{\UnlUser\}} \hat{\vg}_t^u.
\end{align}
\end{enumerate}

\section{Theoretical Evaluation}\label{subsec:analysis}
\ac{medu} can be shown to significantly reduce the memory footprint  via \eqref{eq:unlearned_model}. Still, it  induces  distortion incorporated into the unlearned model. To quantify these aspects, we introduce a set of  assumptions in Subsection~\ref{subsec:analysis_assumption}, under which we characterize storage and distortion in Subsection~\ref{subsec:compression_analysis}, as well as their effect on the unlearned model in Subsection~\ref{subsec:convergence_bound}, and conclude with a discussion in Subsection~\ref{ssec:discussion}. 

\subsection{Assumptions}\label{subsec:analysis_assumption} 
\subsubsection{Compression Operation} For tractability and completeness of our analysis, we first introduce the following assumptions:
\begin{enumerate}[label={\em AS\arabic*},series=assumptions]
    \item \label{itm:zero_overeloading}
    The quantizer employed in encoding step \ref{itm:E step Quantization} and defined in Def.~\ref{def:quantizer}, is not overloaded; i.e., 
    $\Pr\left[\|\vx+\vd\|<\gamma\right]=1.$
    \item \label{itm:diff_thresh_noise}
    The error induced by the differential thresholding  in \ref{itm:E step Differential Thresholding}, given by
     \(\ve^{\rm thresh}_{t,u,m}:=\vg_{t,m}^u - \hat\vg_{{\tau_u(t)},m}^u \ \forall u,t,m\),
    is modeled as a zero-mean  noise which is uncorrelated between users, i.e., 
    \(\E\left[\langle\ve^{\rm thresh}_{t,u,m},\ve^{\rm thresh}_{t',u',m}\rangle\right] = 0\), and has a decaying correlation for each user, where 
    \(\big|\E\left[\langle\ve^{\rm thresh}_{t,u,m},\ve^{\rm thresh}_{t',u,m}\rangle\right]\big| \leq C/|t-t'|^\zeta\)
    for $u'\neq u$, $t'\neq t$ for some constants $C>0$, $\zeta>1$.
\end{enumerate}
Once \ref{itm:zero_overeloading} is satisfied for probabilistic lossy compression such as \ac{sdq}, its associated distortion can be represented as additive zero-mean noise with bounded variance~\cite{zamir2014lattice}. This is also the case with the differential thresholding operation according to \ref{itm:diff_thresh_noise}, as the variance of this error can be shown to be bounded due to its inherent construction in~\eqref{eqn:DiffThresh}. These properties of the distortion components notably facilitate analyzing their impact on the unlearning procedure in the sequel, where aggregation in \eqref{eq:compressed_unlearned_model2} results in these (independent) additive noises terms effectively approaching their mean value of zero by the law of large numbers. 

\smallskip
\subsubsection{Learning Task} 
We continue with adopting the following assumptions on the clients gradients evaluated at either the \ac{fl} global model $\vw_t$ in~\eqref{eq:FedAvg_update} or the train-from-scratch  $\vw^\star_t$ in~\eqref{eq:train_from_scratch}: 
\begin{enumerate}[label={\em AS\arabic*},resume*=assumptions]    
    \item \label{itm:full_gradient_decay}
    The (full) loss gradient decays over time, i.e., there exist positive constants $A,\nu,\alpha$ such   that
    $$\|\nabla \cL_u(\vw^\star_t)\|^2,\|\nabla \cL_u(\vw_t)\|^2\leq \frac{A}{(j+\nu)^{\alpha}}.$$
    \item \label{itm:correlation_decay}
    The correlation of the stochastic gradients from different rounds decays as they grow apart, i.e., 
    for constants $B>0$, $\beta>1$ and any $t'\neq t$ it holds that
    \begin{align*}
    \left|\E\left[\left\langle\nabla \cL_u(\vw_t^1; \xi^u_t) , \nabla \cL_u(\vw_{t'}^2; \xi^u_{t'})\right\rangle\right]\right|
    \leq \frac{B}{|t-t'|^\beta};
    \end{align*}
    for $\vw_t^1,\vw_{t'}^2$ being either $\vw_t,\vw_{t'}; \ \vw^\star_t,\vw^\star_{t'};$ or $\ \vw^\star_t,\vw_{t'}$.
    \item \label{itm:bounded_norm}    
     The expected squared norm of the stochastic gradient is uniformly bounded, i.e.,
    $$\E\left[\|\nabla \cL_u(\vw_t; \xi^u_t)\|^2\right]
    =\E\left[\|\vg_t^u\|^2\right]\leq G^2.$$
\end{enumerate}

To highlight the relevance of the above assumption, we recall that $\vw^\star_t$ differs from $\vw_t$ in having the former representing the retrained \ac{fl} model ({\em without} the $\UnlUser$th user participation), and the latter is the \ac{fl} model learned {\em with} the $\UnlUser$th user. Therefore, it is assumed that  \ac{fl} converges to the optimal model $\vw^\star$ in \eqref{eq:FL_optimal_model} (see \cite{li2019convergence}); namely,
\(\vw^\star_t, \vw_t{\to} \vw^\star\) as $t \to\infty$. In such commonly assumed settings,  the users local updates (gradients evaluated at the global model) decays over time, and thus \ref{itm:full_gradient_decay} is likely to hold.

To understand \ref{itm:correlation_decay}, we note that stochastic gradients taken at different rounds of the \ac{fl} or train-from-scratch procedures may exhibit some level of correlation, due to the learning recurrence formulate. Nevertheless, according to \ref{itm:correlation_decay}, and in line with \ref{itm:diff_thresh_noise}-\ref{itm:full_gradient_decay}, it is assumed that once the model moves towards a stationary point, gradients from different rounds tend to correlate less.
Finally, \ref{itm:bounded_norm} is commonly adopted in convergence studies of distributed learning \cite{li2019convergence, stich2018local, koloskova2019decentralized}. 

\vspace{-0.1cm}
\subsection{Compression Analysis}\label{subsec:compression_analysis} 
\subsubsection{Storage Reduction} 
First, recall that the users updates encoding steps \ref{itm:E step User Selection}-\ref{itm:E step Quantization} involve 
a random user selection $[U]\mapsto \bar{\cU}_t$. Then, the updates of each are divided into sub-vectors, with only the ones not being differentially-thresholded are in $\bar{\cM}_t^u$ \eqref{eq:indices directly stored}, and therefore quantized via \eqref{eq:lattice_quantizer}. As a result, encoding each of the $M/L$ sub-vectors requires storing $1$ bit, deferring whether it is in $\bar{\cM}_t^u$, and in case it does, additional $\log_2(|\cP|)$ bits are needed due to quantization. The overall number of bits stored is thus
\begin{align}
   \!\! \boxed{\text{Storage}_{\rm MEDU}\!= \!
    \sum_{t=0}^{T} \sum_{u\in \bar{\cU}_t}\!\left(\!\frac{M}{L} \!+\! |\bar{\cM}_t^u| \log_2(|\cP|)\!\right)\text{[bits]}.}
    \label{eqn:StorageComp}
\end{align}

As opposed to the conventional memory footprint in \eqref{eqn:StorageFU}, which is fixed and deterministic, the storage in \eqref{eqn:StorageComp} is random (via $\bar{\cU}_t$) and input varying (via $\bar{\cM}_t^u$). Still, we can bound it as follows
\begin{proposition}\label{lem:storage}
    The overall bits in \eqref{eqn:StorageComp} hold (w.p. 1)
    \begin{align}\label{eqn:storage}
        {\rm Storage}_{\rm MEDU} &\leq  \frac{\bar{U}}{U} \cdot 
        \frac{\left(1+\log_2(|\cP|)\right)/L}{b}
        \cdot {\rm Storage}_{\rm FU}.         
    \end{align}
\end{proposition}
\begin{IEEEproof}
    The proposition follows as $|\bar{\cM}_t^u| \leq M/L$ by \eqref{eq:indices directly stored}.
\end{IEEEproof}
\smallskip

Proposition~\ref{lem:storage} indicates that \ac{medu} allow for memory savings by at least a factor which decays with the quantization resolution ($|\cP|$) and number of selected users ($\bar{U}$). These are translated into substantial storage reduction, as we  show in Section~\ref{sec:experiments}. 

\smallskip
\subsubsection{Induced Distortion} 
We next  analyze the compression error. This error is dictated by the distance between the non-compressed unlearned model in~\eqref{eq:unlearned_model} to its counterpart recovered from compressed users updates, given in~\eqref{eq:compressed_unlearned_model} where the recovered gradients of \ac{medu} are obtained from~\eqref{eq:compressed_unlearned_model2}.

With \ref{itm:zero_overeloading}-\ref{itm:diff_thresh_noise}, \ac{medu} includes three random components: the compression distortion, the user selection randomness in encoding step \ref{itm:E step User Selection}, and the \ac{sgd} sample index $\xi^u_t$ in \eqref{eq:lsgd}. This  implies that the compressed unlearned model $\vw''_t$ is random vector, and that  the first- and second-moments of the compression error (compared with the non-compressed counterpart) obey:

\begin{theorem}\label{thm:compression_error_moments}
When \ref{itm:zero_overeloading}-\ref{itm:diff_thresh_noise} and \ref{itm:correlation_decay}-\ref{itm:bounded_norm} hold, $\vw''_{T+1}$  is an unbiased estimator of the non-compressed $\vw'_{T+1}$ in~\eqref{eq:unlearned_model}, i.e.,
${\rm Bias}^{(T)}_{\rm MEDU}=\E\left[\vw''_{T+1} -\vw'_{T+1}\right]=0$;
with variance
\begin{align}\label{eq:compression_error_momentII}
    &{\rm Var}^{(T)}_{\rm MEDU}=\E\left[\left\|\vw''_{T+1} -\vw'_{T+1} \right\|^2\right] \notag  \\
    &\leq2S\Bigg(G^2\sum_{t=0}^{T} \eta^2_t+ B\sum_{\substack{t,t'=0,0\\ t\neq t'}}^{T,T}  \frac{\eta_t \eta_{t'}}{|t-t'|^\beta}\Bigg) +
    \left(S+\frac{1}{U-1} \right)\times \notag\\
    &\qquad \frac{4M}{L}\Bigg(\sum_{t=0}^{T} \eta^2_t(\delta_t^2+\sigma^2_{\rm SDQ})+C\sum_{\substack{t,t'=0,0\\ t\neq t'}}^{T,T}  
    \frac{\eta_t \eta_{t'}}{|t-t'|^\zeta}\Bigg).    
\end{align}
In \eqref{eq:compression_error_momentII}, $\sigma^2_{\rm SDQ}$ is the fixed second-moment of the \ac{sdq} distortion while $S$ is being set according to the sampling scheme of \ref{itm:E step User Selection}, i.e.,
\begin{equation*}    
S:=\frac{1}{\bar{U}-1}\begin{cases}
    1 & \text{clients are sampled with replacement}\\
    \frac{U-\bar{U}}{U-2} & \text{otherwise.}
    \end{cases}    
\end{equation*}
\end{theorem}
\begin{IEEEproof}
    The proof is given in Appendix~\ref{app:compression_error_moments_proof}. 
\end{IEEEproof}
\smallskip
Note that the distortion induced specifically by quantization does not explicitly depend on the number of used bits. This  is encapsulated in the second-order moment $\sigma^2_{\rm SDQ}$ in \eqref{eq:compression_error_momentII}. 
Not surprisingly, it is further revealed that the variance associated with \ac{medu}'s error decreases when: $(i)$ increasing the total amount of \ac{fl} users (high $U$); $(ii)$ sampling a random subset of client {\em with} replacement (low $S$); and/or $(iii)$ quantizing multiple entries of high-dimensional vectors together (high $L$).
The controllable statistical properties associated with \ac{medu}, presented in Lemma~\ref{thm:compression_error_moments},   allow to analytically derive an upper bound on the unlearning procedure convergence, detailed next. 

\subsection{Convergence Bound}\label{subsec:convergence_bound} 
We proceed to characterize the proximity of the  unlearned model of \ac{medu} in~\eqref{eq:compressed_unlearned_model2} to the train-from-scratch one in~\eqref{eq:train_from_scratch}. This deviation from the desired model is highly adopted as a conventional metric in unlearning literature \cite{wang2024machine, wu2020deltagrad, cao2023fedrecover}.
In light of Lemma~\ref{thm:compression_error_moments} in particular and the additive-noise nature of the distortion of \ac{medu}, analyzing the deviation also specializes non-compressed distributed unlearning \eqref{eq:unlearned_model} as a private case, as
\begin{align}\label{eq:sum_of_differences}
    \vw^\star_{T+1} \!-\! \vw''_{T+1} \!=\! 
\underbrace{(\vw^\star_{T+1}\!-\! \vw'_{T+1})}_{\text{non-comp. unlearning error}}\! + \!
\underbrace{(\vw'_{T+1}\!- \!\vw''_{T+1})}_{\text{compression error}}.
\end{align}
Observing this distance as a sum of differences allows to isolate and quantify the residual compression distortion added to the more-accurate-yet-storage-demanding non-compressed  alternative. 
Accordingly,  we commence with the special case of non-compressed server-side unlearning, and extend it to \ac{medu}. 

\smallskip
\subsubsection{Non-Compressed Analysis} 
The assumptions introduced in Subsection~\ref{subsec:analysis_assumption} allow us to rigorously characterize the first two moments of non-compressed server-side unlearning for distributed \ac{sgd}-based learning, with respect to the desired model. This is stated in the following theorem: 
\begin{theorem}\label{thm:moments_without_comp}
When \ref{itm:full_gradient_decay}-\ref{itm:bounded_norm} hold, the expected deviation (moment I) from   the 
train-from-scratch model $\vw^\star_{T+1}$ in~\eqref{eq:train_from_scratch} to the
(non-compressed) unlearned one $\vw'_{T+1}$ in~\eqref{eq:unlearned_model}, $ {\rm MI}^{(T)}_{\rm FU}:=
     \E\left[\vw^\star_{T+1} - \vw'_{T+1}\right]$, equals
\begin{align}
     {\rm MI}^{(T)}_{\rm FU}&=
     \sum_{t=0}^T\eta_t
    \frac{1}{U-1}
    \sum_{\substack{u=1\\ u \neq \UnlUser}}^U      
    \left( \nabla \cL_u(\vw^\star_t) - \nabla \cL_u(\vw_t)\right),\label{eq:non_comp_moment1}
\end{align}
while  $ {\rm MII}^{(T)}_{\rm FU}:=
    \E\left[\left\|\vw^\star_{T+1} - \vw'_{T+1}\right\|^2 \right]$ satisfies
\begin{align}
    {\rm MII}^{(T)}_{\rm FU}
    &\leq4G^2\sum_{t=0}^T\eta^2_t +
    4A^2
    \sum_{t=0}^T \frac{\eta_t}{(t+\nu)^{\alpha}}
    \sum_{\substack{t'=0\\ t'\neq t}}^T 
    \frac{\eta_t\eta_{t'}}{(t'+\nu)^{\alpha}} \notag\\
     \quad&+ 
    \frac{4B}{U-1}\sum_{\substack{t,t'=0,0\\ t\neq t'}}^{T,T}   
    \frac{\eta_t\eta_{t'}}{|t-t'|^\beta}\label{eq:non_comp_moment2}.
\end{align}
\end{theorem}
\begin{IEEEproof}
    The proof is given in Appendix~\ref{app:moments_without_comp_proof}. 
\end{IEEEproof}
\smallskip
Theorem~\ref{thm:moments_without_comp}, and particularly \eqref{eq:non_comp_moment2},  imply that standard server-side unlearning via Algorithm~\ref{alg:fu} does not guarantee a decaying distance from the desired train-from-scratch model as the number of global iterations $T$ grows. This result is not surprising, as raised from the construction of the unlearned model in~\eqref{eq:unlearned_model}, noticing that the remaining users past gradients $\{\cL_u(\vw_t)\}_{u\neq \UnlUser,t}$ still encapsulate information about the $\UnlUser$th user via $\{\vw_t\}_t$ in \eqref{eq:FedAvg_update}. Still, as we show later in Subsection~\ref{ssec:asymptotic}, the result can serve to select a learning step-size value $\eta_t$ for which the unlearned model can be guaranteed to be within some bounded environment of the desired model. 

\smallskip
\subsubsection{Compressed Analysis} 
While Theorem~\ref{thm:moments_without_comp} considers non-compressed unlearning, combining it with the findings of Lemma~\ref{thm:compression_error_moments} allows us to characterize convergence with our proposed \ac{medu}.  The result is stated in the following theorem:

\begin{theorem}\label{thm:moments_with_comp}
When \ref{itm:zero_overeloading}-\ref{itm:bounded_norm} hold, the first two moments of the proximity of the 
train-from-scratch model $\vw^\star_{T+1}$ in~\eqref{eq:train_from_scratch} to the
compressed unlearned one $\vw''_{T+1}$ in~\eqref{eq:compressed_unlearned_model2} satisfy
\begin{align}
     {\rm MI}^{(T)}_{\rm MEDU}&:=
     \E\left[\vw^\star_{T+1} - \vw''_{T+1}\right]=
     {\rm MI}^{(T)}_{\rm FU},\label{eq:comp_moment1}    \\
    {\rm MII}^{(T)}_{\rm MEDU}&:=
    \E\left[\left\|\vw^\star_{T+1} - \vw''_{T+1}\right\|^2 \right] \notag\\
    &\leq
     {\rm MII}^{(T)}_{\rm FU} + {\rm Var}^{(T)}_{\rm MEDU}. \label{eq:comp_moment2}
\end{align}
\end{theorem}
\begin{IEEEproof}
    The theorem follows  from representing $\vw^\star_{T+1} - \vw''_{T+1}$ as in~\eqref{eq:sum_of_differences}, and taking the stochastic expectation. Using the law of total expectations, the terms are shown to be uncorrelated, and  the theorem follows from Theorems \ref{thm:compression_error_moments}-\ref{thm:moments_without_comp}. 
\end{IEEEproof}
\smallskip
Theorem~\ref{thm:moments_with_comp} indicates that convergence with \ac{medu} is tightly bundled with non-compressed  \ac{fu}. This is attributed to the random nature of either of the employed user-selection or quantization. In particular, whereas the integration of the hierarchical lossy compression scheme introduced in Section~\ref{sec:method} does not change the first moment derived in \eqref{eq:non_comp_moment1}, its contribution to the second moment can explicitly be bounded via \eqref{eq:comp_moment2}.

 \smallskip
\subsubsection{Asymptotic Regime} 
\label{ssec:asymptotic}
The deviations of the models $\vw'_{T+1}$ and $\vw''_{T+1}$ from $\vw^\star_{T+1}$ characterized in Theorems~\ref{thm:moments_without_comp}-\ref{thm:moments_with_comp}, respectively, reveal two main properties.
The first is that server-side unlearning, whether non-compressed or using our \ac{medu}, yields a biased estimate of the desired $\vw^\star_{T+1}$. This finding was reported in the unlearning literature, and can be tackled via, e.g., introducing some correction (or skewness) term in unlearning \cite{huynh2024fast}.

The second observation, which is unique to our analysis, is that the deviation, while not diminishing, 
can be made to converge as $T\rightarrow \infty$. Specifically, when the learning rate $\eta_t$ is carefully chosen to gradually not-too-fast decay over time, as highly adopted in other studies of distributed optimization \cite{stich2018local, koloskova2019decentralized}; and was also proved to be a necessity for \ac{fl} convergence \cite[Thm. 4]{li2019convergence} -- we obtain the following proposition regarding the asymptotic behaviors of $\vw'_{T+1}$ and $\vw''_{T+1}$: 

\begin{proposition}\label{lemma:without_comp_asym}
    When \ref{itm:zero_overeloading}-\ref{itm:bounded_norm} hold and the learning rate is set for any $t\geq 0$ as $\eta_t=\frac{a}{t+b},\ a,b>0$; then 
    \begin{subequations}
    \begin{equation}\label{eq:MII_FU_asym}
    {\rm MII}^{(\infty)}_{\rm FU}:=\mathop{\lim}\limits_{T\rightarrow \infty}    
    \E\left[\left\|\vw^\star_{T+1} - \vw'_{T+1}\right\|^2 \right] < \infty,
    \end{equation} 
    \begin{equation}\label{eq:MII_MEDU_asym}
    {\rm Var}^{(\infty)}_{\rm MEDU}:=\mathop{\lim}\limits_{T\rightarrow \infty}    
    \E\left[\left\|\vw''_{T+1} -\vw'_{T+1} \right\|^2\right] < \infty.    
    \end{equation} 
    \end{subequations}
\end{proposition}
\begin{IEEEproof}
    The proof is given in Appendix~\ref{app:double_sum_convergence_proof}. 
\end{IEEEproof}
\smallskip

Proposition~\ref{lemma:without_comp_asym} shows that, when  \ac{sgd}-based \ac{fl} is combined with proper step-size setting, a subsequent unlearning mechanism can obtain a convergent deviation from the desired model, as opposed to a generic divergent bound in \cite{huynh2024fast}. This non-vanishing gap between the non-compressed unlearned model and the train-from-scratch one is also experimentally visualized in the simulations of Section~\ref{sec:experiments}. As raised by \eqref{eq:MII_MEDU_asym}, while \ac{medu} induces additional distortion, its effect does not result in divergence,  as stated in the following corollary. 

\begin{corollary}\label{cor:with_comp_asym}
    When \ref{itm:zero_overeloading}-\ref{itm:bounded_norm} hold and the learning rate is set for any $t\geq 0$ as $\eta_t=\frac{a}{t+b},\ a,b>0$; then it holds that
    \begin{align*}
    {\rm MII}^{(\infty)}_{\rm MEDU}:=\mathop{\lim}\limits_{T\rightarrow \infty}    
    \E\left[
    \left\|\vw^\star_{T+1} - \vw''_{T+1}\right\|^2  
    \right]  < \infty.
    \end{align*}
\end{corollary}
\begin{IEEEproof}
It follows directly from Proposition~\ref{lemma:without_comp_asym} by \eqref{eq:comp_moment2}.
\end{IEEEproof}
\smallskip

Corollary~\ref{cor:with_comp_asym} indicates that while the memory footprint of server-side unlearning can significantly be relieved by \ac{medu}, its excessive deviation due to compression is quantified to be finite for growing number of \ac{fl} global rounds, i.e., $T\to\infty$.

\subsection{Discussion}\label{ssec:discussion}
With the growing popularity of \ac{fl}, the ability to perform server-side unlearning becomes more and more relevant. It allows handling likely situations, where, e.g.,  some user is detected at some point, perhaps even much after training is concluded, as being malicious, compromised, buggy, or just wishes to be forgotten. Accordingly, our proposed \ac{medu} tackle a core obstacle in unlearning, alleviating the bottleneck associated with the immense storage requirements of unlearning via dedicated lossy source coding tools. 
The usage of  a hierarchical lossy compression scheme that exploits the inherent ability to share a source of randomness during compression and recovering,  makes \ac{medu} attractive and easy to apply.

Given the scarcity of convergence characterizations for standard (non-compressed) server-side unlearning, we focus on basic learning-unlearning and compression setups; to have a tractable yet informative analysis of the unlearned model. 
Yet, even with our adopted assumptions, we managed to obtain surprising results. One of which is that our upper bound on the difference between the desired model that is trained from scratch and the model unlearned from lossy compressed stored updates, outperforms the state-of-the-art known bounds for non-compressed decentralized server-side unlearning, even when lossy compression is incorporated. 
Our characterization also specializes a bound for non-compressed distributed unlearning, which was not available to date. 
As an example, \cite{huynh2024fast} only postulates that the growth rate of this difference (between non-compressed unlearning and train-from-scratch models) is at most exponential with the number of \ac{fl} iterations. In our analysis we rigorously characterized the first- and second-order moments of the unlearned model using our proposed \ac{medu}, while also specializing non-compressed settings. 

Nevertheless, it is noted that analytical studies of \ac{fl} \cite{stich2020error} and \ac{fu} \cite{tao2024communication} (though not for server-side settings) exist in literature with relaxed assumptions. Our characterization is therefore expected to motivate future research on analyzing more involved and less restrictive scenarios, as was done in the domain of compressed \ac{fl}. In particular, the derived theoretical analysis may be an important first step toward a proof using relaxed assumptions, e.g., go beyond \ac{sgd}-based learning. 
Additionally, as shown in recent (non server-side) \ac{fu} works, in some scenarios the  unlearned model may actually converge to the optimal one. We thus expect our upper bound to serve as a starting point for proving the convergence of server-side unlearning. 

Beyond extending our theoretical analysis, our work gives rise to numerous  research directions. From an algorithmic standpoint, the combination of compressed gradients can be extended to unlearning a group of users, as well as having additional users aiding the server in unlearning. 
On the compression side, storage can possibly be further reduced  without affecting operation by adding lossless compression, e.g., entropy codes. When such further compression is present, one can employ  entropy-constrained quantization for lossy compression~\cite{chou1989entropy}. 
Another study can extend the  encoding~\ref{itm:E step Quantization} and decoding~\ref{itm:D Quantization Decoding} steps to encompass other probabilistic quantization schemes such as non-subtractive \ac{dq}, where there, the dither is not re-realized in decoding, and can be rendered in encoding such that its associated distortion observed as additive noise; though of larger error \cite{kirac1996results}. 
Finally, additional lead would try to mitigate the overloaded quantizers numerically evidenced in Section~\ref{sec:experiments} by designing them~\cite{shlezinger2019hardware} or their associated lattices~\cite{lang2024data} in a learning-oriented rather than a distortion-oriented manner. These extensions are left for future work.


\begin{figure}
\centering
    \begin{minipage}{0.48\linewidth}
        \centering
        \includegraphics[width=\linewidth]{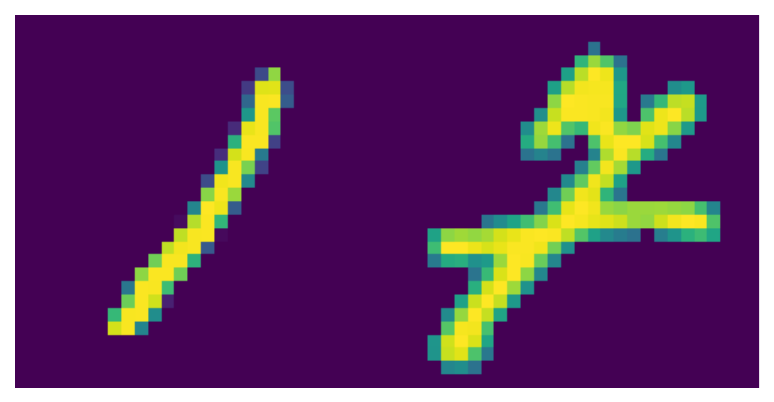}
    \end{minipage}%
    \hfill
    \begin{minipage}{0.48\linewidth}
        \centering
        \includegraphics[width=\linewidth]{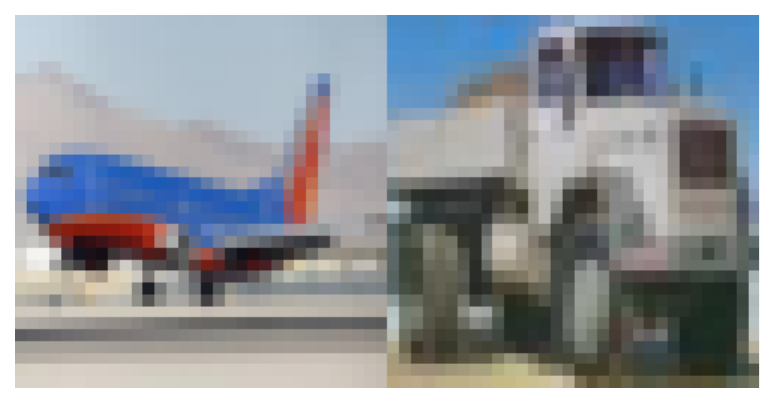}
    \end{minipage}
\caption{Backdoors examples: a digit `$7$' is labeled as `$1$' for MNIST, and an airplane is labeled as `truck' for CIFAR-$10$.}
\label{fig:backdoor_illustration}           
\end{figure}

 \section{Numerical Experiments}\label{sec:experiments}
 We next experimentally evaluate server-side unlearning with \ac{medu}\footnote{The source code used in our experimental study, including all the hyper-parameters, is available online at \url{https://github.com/alonhelvits/FedUL_Quant}.}. We first detail the experimental setup in Subsection~\ref{subsec:setting}, after which we present the results in Subsection~\ref{ssec:Results}.

\subsection{Experimental Setup}\label{subsec:setting}  
\subsubsection{Learning Task} 
We consider \ac{fu} using two image classification datasets of MNIST and CIFAR-$10$. For each, we train a \ac{cnn} composed of two convolutional layers and two fully-connected ones; with intermediate ReLU activations, max-pooling and normalization layers.
The data is  distributed across $U=25$ users, where non-IID division is obtained   by randomizing label division based on the Dirichlet distribution with parameter $\textcolor{red}{???}$ \cite{li2022federated}. Prior to unlearning, \ac{fl} is globally iterated over $T=30$ rounds, each client locally utilizing \ac{sgd} for $10$ epochs.

\subsubsection{Algorithms} 
\label{sssec:benchmarks}
We compare \ac{fu} using \ac{medu} with  three reference baselines, which are 
{\em Original \ac{fl}}, i.e., the converged model $\vw_{T+1}$ of the \ac{fl} training in~\eqref{eq:FedAvg_update}; 
{\em Retrain}, denoting the desired train-from-scratch model $\vw^\star_{T+1}$ in~\eqref{eq:train_from_scratch}; and
{\em non-compressed \ac{fu}}, which uses the unlearned model $\vw'_{t+1}$ obtained via~\eqref{eq:unlearned_model}.  

These are compared to the compressed unlearned model $\vw''_{T+1}$ obtained via~\eqref{eq:compressed_unlearned_model}. To evaluate the effect of each of the components of \ac{medu}, we evaluate different combinations of its parameters. These are determined by the user selection numbers $\bar{U}$ in \ref{itm:E step User Selection} (which is canceled  by setting $\bar{U}=U$); the differential threshold $\tau_u(t)$ in \ref{itm:E step Differential Thresholding} (which is abbreviated here as $\tau_u$ and is canceled for $\tau_u\equiv 0$); and the quantization  resolution $\log_2|P|$ (which is ${64}$  for non-quantized). Specifically, we consider uniform scalar (SQ, $L=1$) and hexagonal lattice (VQ, $L=2$) quantizers.

\subsubsection{Evaluation Metrics} 
For each algorithm, we evaluate three performance measures: $(i)$ Learning performance, representing {\em primary task accuracy} on the test data; $(ii)$ Number of bits stored, i.e., {\em Bit Budget},  where for non-quantized weights are saved at double precision; and $(iii)$ Unlearning accuracy. To assess the latter, we employ the established edge-case backdoor of \cite{wang2020attack}, where an adversary edge device intentionally uses wrong labels for a specific set of data points to mislead the server on seemingly easy inputs. This attack,  illustrated in Fig.~\ref{fig:backdoor_illustration},   allows evaluating the unlearning of the adversarial device via the {\em backdoor accuracy}, for which high values indicate that the unlearned model is still backdoored \cite{wang2024machine}.

\input{Tikz_figs/Ablation_MNIST}

\subsection{Results}
\label{ssec:Results}
\subsubsection{Compression Factors} 
We first evaluate the effect of each of the compression factors included in \ac{medu} on the resulting tradeoff between the key performance metrics for MNIST. In each experiment, we apply Algorithm~\ref{alg:compressed fu} to completely unlearn the contribution of the omitted user, such that all unlearned models achieve backdoor accuracy of $0\%$. Thus, we report solely the (primary task) accuracy versus memory reduction ($\frac{\text{Storage}_{\rm MEDU}}{\text{Storage}_{\rm FU}}$). 
In Fig.~\ref{fig:results}  we evaluate the accuracy and memory reduction values achieved when altering a single compression hyperparameters: the quantization rate (Fig.~\ref{fig:resultsA}); the differential threshold (Fig.~\ref{fig:resultsB}); and the number of selected users (Fig.~\ref{fig:resultsC}).

\input{Tikz_figs/Tradeoff_MNIST}

We observe in Fig.~\ref{fig:resultsA} that when disabling \ref{itm:E step User Selection} ($\bar{U}=25$) and \ref{itm:E step Differential Thresholding} ($\tau_u = 0$), then increasing the quantization rate yields gradual performance improvements, with VQ outperforming SQ at similar rates. However, when including also user selection and differential thresholding, monotonicity no longer holds as the distortions induced by each compression scheme do not necessarily accumulate, and we observe notable memory reductions that do not lead to accuracy degradation. When altering $\tau_u$ in Fig.~\ref{fig:resultsB}, we observe more dramatic variations in the memory reduction compared to changing the quantization rate, with lesser effect on the primary task accuracy, depending on the level of user selection and quantization employed, indicating the usefulness of this compression form in enhancing memory reduction. Finally, we note that the number of users $\bar{U}$ affects the memory reduction roughly as a constant factor (e.g., setting $\bar{U}=15$ induces a factor of $\approx \frac{15}{25}=0.6$,  and that when combined with stochastic quantization and differential thresholding, its effect does not necessarily lead to reduced accuracy.

\input{Tikz_figs/Tradeoff_CIFAR}

\subsubsection{Baselines Comparison}
Having identified the complementary effect of the compression ingredients of \ac{medu}, we next show their resulting tradeoffs in comparison to the benchmarks detailed in Subsection~\ref{sssec:benchmarks}. The different levels of accuracy-memory tradeoffs of \ac{medu} trained for MNIST are illustrated in Fig.~\ref{fig:mnist_tradeoff}. There, we observe that \ac{medu} consistently unlearns the backdoor attack and achieves similar accuracy to non-compressed \ac{fu}, while notably reducing the memory requirements. In particular, for memory utilization larger than $40\%$, \ac{medu} achieves similar accuracy as that of non-compressed \ac{fu}, while supporting enhanced reductions that are gracefully balanced with main tasks  accuracy. It is noted though that in terms of primary task accuracy, server-side unlearning leads to about $11\%$ reduction in primary task accuracy compared to the original \ac{fl} model (which is backdoored) and training from scratch (which necessitates re-gathering all the users and initiating a lengthy \ac{fl} procedure).

The accuracy loss due to server-side unlearning is more dominant for the CIFAR dataset, as reported in Fig.~\ref{fig:cifar_tradeoff}. There, server-side \ac{fu} leads to degradation of over $20\%$ in primary task accuracy. In this case, it is observed that the distortion induced by \ac{medu} can in fact lead to an improved model compared to non-compressed \ac{fu}, while reducing memory footprint by over an order of magnitude. It is also noted that all models have non-zero backdoor accuracy here, as there is a non-negligible probability of misclassifying the images targeted by the backdoor attack (e.g., labeling the selected airplane images as 'truck') even if no attack takes place. Still, unlearning notably reduces the probability of this happening compared to the original \ac{fl} model.

The results presented so far demonstrate that \ac{medu} allows implementing server-side unlearning with notable memory reduction and with minor effect on the primary task accuracy compared to non-compressed \ac{fu}. The performance reduction compared to the non-\ac{fu} models observed in Figs.~\ref{fig:mnist_tradeoff}-\ref{fig:cifar_tradeoff} stem from server-side unlearning degradation, and existing \ac{fu} algorithms typically combine additional training rounds following server-side \ac{fu} to mitigate this effect~\cite{romandini2024federated}. To show that the same approach can be applied with \ac{medu}, we took two instances of \ac{medu} for CIFAR (where a notable degradation compared to non-\ac{fu} is observed) with memory utilization of $2\%$ and $4.9\%$, and carried out additional adaption rounds using merely $5$ users. The results, reported in Table~\ref{tab:further}, demonstrate that the performance gap notably decreases within a few rounds compared to the desired train-from-scratch model, which requires all users to be available and fully restart the whole learning procedure. These results showcase that the memory reduction gains of \ac{medu} can be combined with existing \ac{fu} methods to mitigate performance loss.

\section{Conclusions}\label{sec:conclusions}
In this work, we studied decentralized server-side unlearning with lossy source coding incorporated in the process. We investigated its effect on the unlearned model performance from both theoretical and experimental perspectives. From an experimental point of view, our numerical evaluations reveal that compressed \ac{fu} with notable storage reduction, e.g., $32\times$ lower memory footprint at the server, preserves the ability to unlearn. On the theoretical side; we prove that under common assumptions, the distance between the compressed unlearned model to the desired one is bounded. Our bound improves upon the best known guarantees, asserting that the growth rate of this distance is at most exponential with the number of \ac{fl} iterations. 

\begin{table}[]
\caption{Primary task accuracy with further training.}
\label{tab:further}
 \begin{adjustbox}{width=0.48\textwidth} 
\begin{tabular}{|c|ccc|ccc|c|}
\hline
\rowcolor[HTML]{C0C0C0} 
                                 & \multicolumn{3}{c|}{\cellcolor[HTML]{C0C0C0}MEDU (4.9\%)}                                               & \multicolumn{3}{c|}{\cellcolor[HTML]{C0C0C0}MEDU (2\%)}                                                 & \cellcolor[HTML]{C0C0C0}                                                                                \\ \cline{1-7}
\rowcolor[HTML]{C0C0C0} 
Rounds                   & \multicolumn{1}{c|}{\cellcolor[HTML]{C0C0C0}1} & \multicolumn{1}{c|}{\cellcolor[HTML]{C0C0C0}5} & 10    & \multicolumn{1}{c|}{\cellcolor[HTML]{C0C0C0}1} & \multicolumn{1}{c|}{\cellcolor[HTML]{C0C0C0}5} & 10    & \multirow{-2}{*}{\cellcolor[HTML]{C0C0C0}\begin{tabular}[c]{@{}c@{}}Train-from\\ -scratch\end{tabular}} \\ \hline
\cellcolor[HTML]{C0C0C0}Accuracy & \multicolumn{1}{c|}{49.13}                     & \multicolumn{1}{c|}{62.53}                     & 67.02 & \multicolumn{1}{c|}{49.23}                     & \multicolumn{1}{c|}{63.13}                     & 66.95 & 69.46                                                                                                   \\ \hline
\end{tabular}
\end{adjustbox}
\end{table}

\appendix
\section{Appendix}
\numberwithin{lemma}{subsection} 
\numberwithin{corollary}{subsection} 
\numberwithin{remark}{subsection} 
\numberwithin{equation}{subsection}	

It is first noted that while analyzing $\vw''_{T+1}$ in~\eqref{eq:compressed_unlearned_model2}, there are four different sources of randomness involved: 
$a)$ stochastic gradients;
$b)$ randomly sampled devices;
$c)$ probabilistic quantization; and
$d)$ random modeling of differential thresholding,
while either of 
$\vw_t$, $\vw^\star_t$, $\vw'_t$ in \eqref{eq:FedAvg_update},  \eqref{eq:train_from_scratch}, \eqref{eq:unlearned_model}, respectively, only constitutes the factor $a)$. 
As this $a)$-$d)$ quadruplet is mutually independent with respect to both $u$ and $t$; the law of total expectation can be employed in order to isolate each stochastic term. Equivalently, each of which can be distinguished using the notation $\E_{\{\cdot\}}(\cdot)$. Yet, this subscript is dropped in the sequel for ease of notation.

\subsection{Proof of Theorem~\ref{thm:compression_error_moments}}\label{app:compression_error_moments_proof}

\subsubsection{Auxiliary Results}
To convey the proof of Theorem~\ref{thm:compression_error_moments}, 
we begin by stating \cite[Thm. 1]{gray1993dithered} after which it would be necessary to prove certain auxiliary lemmas. We defer the proofs of these lemmas to Appendix~\ref{app:Proofs of auxiliary lemmas}.
\begin{theorem}[\ac{sdq}]\label{thm:SDQ}  
Consider input $\{\vx_i\}$ and dither $\{\vd_i\}$ sequences, where each of the latter is uniformly distributed over the basic cell of a $\gamma$-supported lattice $\cP$; such that $\forall i \ \Pr(\|\vx+\vd\|<\gamma)=1$.
Then, the \ac{sdq} errors, defined $\forall i$ as 
\begin{equation}\label{eq:sdq_error}
    \ve_i^{\rm SDQ} :=  \argmin_{\vl\in\cP} \|\vx_i+\vd_i - \vl\| -\vd_i - \vx_i,
\end{equation}
are mutually independent of $\{\vx_i\}$; and obey an i.i.d uniform distribution over the basic lattice cell. That is, $\forall i$ 
\begin{align}\label{eq:sdq_moments}
    \E\left[\ve_i^{\rm SDQ}\right]=0; \
\Var\left(\ve_i^{\rm SDQ}\right)
=\sigma^2_{\rm SDQ}.
\end{align}
\end{theorem}

\begin{lemma}[Sample Mean]\label{lemma: sample mean}
Let $X_1,X_2,\dots,X_n$ be uniformly at random sampled from a finite population of size $N$ with mean $\mu$ and variance $\sigma^2$; and define the sample mean as $\bar{X}:=\frac{1}{n}\sum_{i=1}^n X_i$. Then, the sample mean is an unbiased estimator of the population mean, i.e., $\E\left[\bar X\right] = \mu$, with variance
\begin{equation}\label{eq:app_sample_mean_var}
\Var(\bar X)=\sigma^2\times
    \begin{cases}
    1/n & \text{sampling with replacement,}\\
    \frac{1}{n}\frac{N-n}{N-1} & \text{otherwise.}
    \end{cases}
\end{equation}
\end{lemma}

\begin{observation}[Squared Norm Sum Inequality]\label{obs:Squared-Norm Inequality}
Let $\{\vx_i\}_{i=1}^n$ be a sequence of $n$ vectors, each of size $n$. Then, by applying $n$ times Cauchy-Schwartz inequality (with the all-ones vector), it holds that  
   $ \big\|\sum_{i=1}^n \vx_i\big\|^2 
    \leq n\sum_{i=1}^n \left\|\vx_i\right\|^2$.
\end{observation}

\begin{lemma}[MEDU's Error Moments]\label{lemma: reconstruction error}
If \ref{itm:zero_overeloading}-\ref{itm:diff_thresh_noise} hold, then \ac{medu}'s reconstruction error, defined $\forall u,t,m$ as 
$\ve^{\rm MEDU}_{u,t,m}:= \hat{\vg}_{t,m}^u - \vg_{t,m}^u$, is of zero-mean 
$\E\left[\ve^{\rm MEDU}_{u,t,m}\right]=0$, at most decaying covariance, and bounded variance. That is, for $u',t', C>0, \zeta>1$ and $\sigma^2_{\rm SDQ}$ in~\eqref{eq:sdq_moments}, it holds that
\begin{align*}
    \Cov\left(\ve^{\rm MEDU}_{u,t,m},\ve^{\rm MEDU}_{u',t',m}\right):=\E\left[{\ve^{\rm MEDU}_{u,t,m}}^T\ve^{\rm MEDU}_{u',t',m}\right]\\=
    \begin{cases}
        0 & u=u',\\
        \leq C/|t'-t|^\zeta & \text{otherwise};
    \end{cases}\\
    \Var\left(\ve^{\rm MEDU}_{u,t,m}\right):= \E\left[\left\|\ve^{\rm MEDU}_{u,t,m}\right\|^2\right]\leq \delta_t^2 + \sigma^2_{\rm SDQ}.
\end{align*}
\end{lemma}

\subsubsection{Proving Theorem~\ref{thm:compression_error_moments}}
Using the explicit definitions of $\vw''_{T+1}, \vw'_{T+1}$ in \eqref{eq:compressed_unlearned_model2}, \eqref{eq:unlearned_model}, respectively; we obtain
    $\E\left[\vw''_{T+1} -\vw'_{T+1}\right] \overset{(a)}{=}
    \sum_{t=0}^{T} \eta_t\frac{1}{U-1} \sum_{\substack{u=1\\ u \neq \UnlUser}}^U 
    \E\left[\hat\vg_t^u - \vg_t^u\right]  =
    \sum_{t=0}^{T} \eta_t\frac{1}{U-1} \sum_{\substack{u=1\\ u \neq \UnlUser}}^U 
    \sum_{m=1}^{M/L}\E\left[\hat\vg_{t,m}^u - \vg_{t,m}^u\right]
    \overset{(b)}{=}0$.
Here, $(a)$ holds by Lemma~\ref{lemma: sample mean}, as $\bar{\cU}_t$ is uniformly sampled out of $[U]$; and $(b)$ follows from Lemma~\ref{lemma: reconstruction error}. 

As for the variance, by defining $\cV_t := \bar{\cU}_t / \{\UnlUser\}$, we have
\begin{align}
    &\E\left[\left\|\vw''_{T+1} -\vw'_{T+1}\right\|^2\right] =\sum_{m=1}^{M/L} \times\notag\\
    &\E\Big[\big\|\sum_{t=0}^{T} \eta_t\big(\frac{1}{|\cV_t| } 
    \!\!\!\sum_{u \in \cV_t} 
    \hat{\vg}_{t,m}^u -\frac{1}{U-1} \sum_{u\neq \UnlUser} \vg_{t,m}^u\big)\big\|^2\Big] \label{eq:inner_summand},
\end{align}
where each summand can then be bounded according to
\begin{align}
    &\eqref{eq:inner_summand}
    \overset{(a)}{\leq}
    2\E\Big[\big\|\sum_{t=0}^{T} \eta_t\frac{1}{|\cV_t | } 
    \!\!\!\sum_{u \in \cV_t }  
    \ve^{\rm MEDU}_{u,t,m}\big\|^2\Big]\label{eq:first_term}\\
    &+
    2\E\Big[\big\|\sum_{t=0}^{T} \eta_t\big(\frac{1}{|\cV_t | } 
    \!\sum_{u \in \cV_t } 
    {\vg}_{t,m}^u\!-\! \frac{1}{U\!-\!1} 
    \sum_{u\neq \UnlUser} \vg_{t,m}^u\big)\big\|^2\Big], \label{eq:second_term} 
\end{align}
where $(a)$ holds by using \ac{medu}'s error definition in Lemma~\ref{lemma: reconstruction error}, i.e.
$\hat{\vg}_{t,m}^u=\ve^{\rm MEDU}_{u,t,m}+\vg_{t,m}^u$; and by Observation~\ref{obs:Squared-Norm Inequality}. 
Now, we bound each term separately. 
Consider Lemma~\ref{lemma: sample mean} while replacing $N=U-1$ and 
$n=|\bar{\cU}_t / \{\UnlUser\}|$, where $|\bar{\cU}_t|=\bar{U}$, then
\begin{align*}
S_t:=\frac{1}{|\bar{\cU}_t / \{\UnlUser\}|}
\begin{cases}
    1 & \bar{\cU}_t \text{ sampled with replace.,}\\
    \frac{U-1-|\bar{\cU}_t / \{\UnlUser\}|}{U-2} & \text{otherwise;}
    \end{cases}    \\ \leq
    \frac{1}{|\bar{\cU}_t|-1}
    \begin{cases}
    1 & \bar{\cU}_t \text{  sampled with replace.,}\\
    \frac{U-1-(|\bar{\cU}_t|-1)}{U-2} & \text{otherwise;}
    \end{cases} \\=
    \frac{1}{\bar{U}-1}
    \begin{cases}
    1 & \bar{\cU}_t \text{ sampled with replace.,}\\
    \frac{U-\bar{U}}{U-2} & \text{otherwise;}
    \end{cases} :=   S.
\end{align*}
Therefore, we get that $\sum_{m=1}^{M/L}\eqref{eq:second_term}$ is upper bounded by
\begin{align}
   &2S\sum_{m=1}^{M/L} \E\Big[\frac{1}{U-1} \sum_{u\neq \UnlUser}\Big\|\sum_{t=0}^{T} \eta_t{\vg}_{t,m}^u-\frac{1}{U-1} \sum_{u\neq \UnlUser} \sum_{t=0}^{T} \eta_t\vg_{t,m}^u\Big\|^2\Big] \notag \\
   &\overset{(a)}{\leq}
    2S\sum_{m=1}^{M/L}\E\Big[\frac{1}{U-1} \sum_{u\neq \UnlUser}\Big\|\sum_{t=0}^{T} \eta_t{\vg}_{t,m}^u\Big\|^2\Big]  =  \frac{2S}{U-1}\sum_{u\neq \UnlUser} \sum_{m=1}^{M/L}\times \notag \\
    &
    \Big(\sum_{t=0}^{T} \eta^2_t\E\left[\|{\vg}_{t,m}^u\|^2\right]+
    \sum_{t\neq t'}  \eta_t \eta_{t'}\E\left[{{\vg}_{t,m}^u}^T{\vg}_{t',m}^u\right]\Big)\label{eq:second_term_intermidiate_bound}\\
    &\overset{(b)}{\leq}
    2S\left(G^2\sum_{t=0}^{T} \eta^2_t+ B\sum_{t\neq t'}  \frac{\eta_t \eta_{t'}}{|t-t'|^\beta}\right).\label{eq:second_term_bound}
\end{align}
In $(a)$ we use $\E\left[\|X-\E\left[X\right]\|^2\right]\leq\E\left[\|X\|^2\right]$ where $X=\sum_{t=0}^{T} \eta_t\vg_{t,m}^u$ w.p. $1/(U-1)$, and $(b)$ holds by \ref{itm:correlation_decay}-\ref{itm:bounded_norm}.

We further bound \eqref{eq:first_term}, 
\begin{align} 
&\hspace{-0.2cm}  \frac{\eqref{eq:first_term}}{2}
    \overset{(a)}{\leq} 2\E\Big[\Big\|\sum_{t=0}^{T} \eta_t\frac{1}{U-1} \sum_{u \neq \UnlUser} \ve^{\rm MEDU}_{u,t,m}
    \Big\|^2\Big] \notag\\
    &+2 \E\Big[\Big\|\sum_{t=0}^{T} \big(\frac{\eta_t}{|\cV_t| } 
    \!\!\!\sum_{u \in \cV_t}  
    \ve^{\rm MEDU}_{u,t,m} - \frac{\eta_t}{U-1} \sum_{u \neq \UnlUser} \ve^{\rm MEDU}_{u,t,m}\big)\Big\|^2\Big]\notag\\
    & 
\hspace{-0.2cm} \overset{(b)}{\leq}\!
    2\big(S \!+\! \frac{1}{U\!-\! 1} \big)\Big(\sum_{t=0}^{T} \eta^2_t(\delta_t^2\!+\! \sigma^2_{\rm SDQ})\!+\!C\sum_{t\neq t'}   
    \frac{\eta_t \eta_{t'}}{|t\!-\! t'|^\zeta}\Big).\label{eq:first_term_bound}
\end{align}
Here $(a)$ follows by adding and subtracting the (all-users) term $\sum_{t=0}^{T} \eta_t\frac{1}{U-1} \sum_{u\neq \UnlUser} \ve^{\rm MEDU}_{u,t,m}$ and using Observation~\ref{obs:Squared-Norm Inequality};
$(b)$ holds as $\E\left[{\ve^{\rm MEDU}_{u,t,m}}^T\ve^{\rm MEDU}_{u,t',m}\right]=0$ according to Lemma~\ref{lemma: reconstruction error}, by repeating the same steps used to derive \eqref{eq:second_term_intermidiate_bound} while substituting $\vg^u_{t,m}$ with $\ve^{\rm MEDU}_{u,t,m}$; and 
using the moments stated in Lemma~\ref{lemma: reconstruction error}.

Altogether, by \eqref{eq:first_term_bound} and \eqref{eq:second_term_bound} we desirably get \eqref{eq:compression_error_momentII}. 

\subsection{Proof of Theorem~\ref{thm:moments_without_comp}}\label{app:moments_without_comp_proof}
To prove Theorem~\ref{thm:moments_without_comp}, first note that due to the uniform sampling of the data samples, 
the following holds,

\begin{observation}\label{obs:stoch_grad_unbiasedness}
    The stochastic gradient is an unbiased estimator of the full gradient, i.e., 
    $\E\left[\nabla \cL_u(\vw_t; \xi^u_t)\right]=\nabla \cL_u(\vw_t)$.
\end{observation}

Using Observation~\ref{obs:stoch_grad_unbiasedness} and the definitions of the train-from-scratch model $\vw^\star_{T+1}$ in~\eqref{eq:train_from_scratch} and the (non-compressed) unlearned one $\vw'_{T+1}$ in~\eqref{eq:unlearned_model}, we desirably obtain that ${\rm MI}^{(T)}_{\rm FU}$ equals
\begin{align*}
     &\E\Big[\sum_{t=0}^{T}\eta_t 
     \underbrace{\frac{1}{U-1}\sum_{\substack{u=1\\ u \neq \UnlUser}}^U     
    \underbrace{\left( \nabla \cL_u(\vw^\star_t; \xi^u_t) - \nabla \cL_u(\vw_t; \xi^u_t) \right)}_{:=B_t^u}}_{:=A_t}\Big] 
    \\
    & \overset{(a)}{=}
   \sum_{t=0}^{T}\eta_t \frac{1}{U-1}\sum_{\substack{u=1\\ u \neq \UnlUser}}^U        
    \left( \nabla \cL_u(\vw^\star_t) - \nabla\cL_u(\vw_t) \right). \notag
\end{align*}

To derive the second-order moment, we utilize the auxiliary definition of $A_t$, noting that ${\rm MII}^{(T)}_{\rm FU} = \E\big[\big\|\sum_{t=0}^{T}\eta_t A_t \big\|^2\big]$, thus
\begin{align}
   \!\!\!\! {\rm MII}^{(T)}_{\rm FU} \!=\!
    \sum_{t=0}^{T}\eta^2_t\E\big[\big\| A_t \big\|^2\big] \!+\!
    \sum_{t=0}^{T}\sum_{\substack{t'=0\\ t'\neq t}}^T
    \eta_t\eta_{t'}\E\left[\left\langle A_t,A_{t'} \right\rangle\right].\label{eq:variance_bound}
\end{align}
We next analyze each summand separately, with  
\vspace{-0.1cm}
\begin{align}\label{eq:A_t non-crossing terms bound}
   & \E\big[\big\| A_t \big\|^2\big] \!=\! 
    \frac{1}{(U-1)^2} 
    \E\Big[\Big\| 
    \sum_{\substack{u=1\\ u \neq \UnlUser}}^U      
   \nabla \cL_u(\vw^\star_t; \xi^u_t) \!-\! \nabla \cL_u(\vw_t; \xi^u_t)  \Big\|^2\Big] \notag\\
    &\overset{(a)}{\leq} 
    \frac{U\!-\!1}{(U\!-\! 1)^2}\sum_{\substack{u=1\\ u \neq \UnlUser}}^U  
    \E\left[\left\|     
    \nabla \cL_u(\vw^\star_t; \xi^u_t) \!-\! \nabla \cL_u(\vw_t; \xi^u_t) \right\|^2\right] 
   \! \overset{(b)}{\leq} \!
    4G^2,
    \vspace{-0.1cm}
\end{align}
where $(a)$ holds by applying Observation~\ref{obs:Squared-Norm Inequality} with $n=U-1$, and so does $(b)$ for $n=2$, while further incorporating \ref{itm:bounded_norm}. For the crossing terms, where $t\neq t'$, we leverage the auxiliary definition of $B^u_t$, as $ \E\left[\left\langle A_t,A_{t'} \right\rangle\right]  :=
    \frac{1}{(U-1)^2} \sum_{\substack{u=1\\ u \neq \UnlUser}}^U \sum_{\substack{u'=1\\ u' \neq \UnlUser}}^U      
    \E\bigl[\bigl\langle B^u_t,B^{u'}_{t'} \bigr\rangle\bigr] $ which equals
    \vspace{-0.1cm}
\begin{align}\label{eq:A_t crossing terms bound}
    \frac{1}{(U-1)^2} \sum_{\substack{u=1\\ u \neq \UnlUser}}^U  
    \Big(\E\bigl[\bigl\langle B^u_t,B^u_{t'}  \bigr\rangle\bigr] + 
    \sum_{\substack{u'=1\\ u' \neq u}}^U   
    \E\bigl[\bigl\langle B^u_t,B^{u'}_{t'} \bigr\rangle\bigr] \Big).
    \vspace{-0.1cm}
\end{align}
The non-crossing terms ($u'=u$) 
      are bounded via \ref{itm:correlation_decay} as
\begin{align}\label{eq:B_t non-crossing terms bound}
\E\bigl[\bigl\langle B^u_t,B^u_{t'} \bigr\rangle\bigr] 
      \leq 4\frac{B}{|t-{t'}|^\beta}. 
\end{align} 
Finally, by applying Observation~\ref{obs:stoch_grad_unbiasedness} and subsequently Cushy-Schwartz inequality, the crossing terms in~\eqref{eq:A_t crossing terms bound} for which $u\neq u'$, hold $\E\bigl[\bigl\langle B^u_t,B^{u'}_{t'} \bigr\rangle\bigr] {\leq} 
      \|\nabla \cL_u(\vw^\star_t)\|\|\nabla \cL_v(\vw^\star_{t'})\|
      +\| \nabla \cL_u(\vw_t)\| \|\nabla \cL_{u'}(\vw^\star_{t'})\|
     +\|\nabla \cL_u(\vw^\star_t)\| \|\nabla \cL_{u'}(\vw_{t'})\| 
      +\|\nabla \cL_u(\vw_t)\| \|\nabla \cL_{u'}(\vw_{t'})\| $, and thus, by \ref{itm:full_gradient_decay}
\begin{align}
\E\bigl[\bigl\langle B^u_t,B^{u'}_{t'} \bigr\rangle\bigr] 
      {\leq}4\frac{A}{(t+\nu)^{\alpha}}\cdot\frac{A}{({t'}+\nu)^{\alpha}}.\label{eq:B_t crossing terms bound}
\end{align}

Altogether, integrating \eqref{eq:B_t non-crossing terms bound}-\eqref{eq:B_t crossing terms bound} into \eqref{eq:A_t crossing terms bound}, and then incorporating \eqref{eq:A_t non-crossing terms bound}-\eqref{eq:A_t crossing terms bound} into \eqref{eq:variance_bound}; the proof of \eqref{eq:non_comp_moment2} is concluded. 

\subsection{Proof of Proposition~\ref{lemma:without_comp_asym}}\label{app:double_sum_convergence_proof}
Note that~\eqref{eq:MII_FU_asym} follows by calculating the limit the  bound  in~\eqref{eq:non_comp_moment2}. There, the first summand  converges for the specified learning rate, and so does the second summand as $\alpha>0$ by \ref{itm:full_gradient_decay}. 
Thus, we next show that the third summand in~\eqref{eq:non_comp_moment2} converges, and the  same argument can be used to derive~\eqref{eq:MII_MEDU_asym} based on~\eqref{eq:compression_error_momentII}.

For $\eta_t=\frac{a}{t+b},\ a,b>0$; we first decompose the double sum such that the lower running indexes would be $1$ instead of $0$, i.e.,
\begin{align}
  &  \sum_{t=0}^{T}\sum_{\substack{t'=0\\ t'\neq t}}^T
    \frac{\eta_t\eta_{t'}}{|t-t'|^\beta} = 
    a^2\sum_{t=0}^{T}
    \sum_{\substack{t=0\\ t'\neq t}}^T    
    \frac{1}{(t+b)(t'+b)|t-t'|^\beta}   \notag\\
    &\leq 
    a^2 \Big(\sum_{t'=1}^T \frac{1/b}{{t'}^{1+\beta}}
    \!+\! \sum_{t=1}^{T}\frac{1/b}{t^{1+\beta}}\!+\! 
     \sum_{t=1}^{T}\sum_{\substack{t'=1\\ t'\neq t}}^T \frac{1}{t\cdot t'|t\!-\! t'|^\beta} \Big). \label{eq:double_sum_decomposition}
\end{align}
In \eqref{eq:double_sum_decomposition}, as $\beta$ is strictly positive, the first two summands converge. To prove that so does the last term,  we utilize the symmetry of $t\cdot t'|t-t'|^\beta$ with respect to both $t$ and $t'$;
we can drop the absolute value and count each value twice, i.e.,  
\begin{align}\label{eq:third_summand}
\sum_{t=1}^{T} \sum_{\substack{t'=1\\ t'\neq t}}^T
    \frac{1}{t\cdot t'|t-t'|^\beta}=
    2\sum_{t=1}^{T} \sum_{t'=t+1}^T
    \frac{1}{t\cdot t'(t'-t)^\beta}\overset{(a)}{\leq}\notag\\
    2\sum_{t=1}^{T} \frac{1}{t^2}
    \sum_{t'=t+1}^T
    \frac{1}{(t'-t)^\beta} 
    \overset{(b)}{\leq}
    2\sum_{t=1}^{T} \frac{1}{t^2}
    \sum_{u=1}^{T}
    \frac{1}{u^\beta}, 
\end{align}
where $(a)$ follows as $t'>t$; in $(b)$ we substitute $u:=t'-t$ and extending the sum's upper limit from $T-t$ to $T$. In the end, \eqref{eq:third_summand} concludes the proof since it is given that $\beta>1$.

\subsection{Proofs of Auxiliary Lemmas}\label{app:Proofs of auxiliary lemmas}
\subsubsection{Proof of Lemma~\ref{lemma: sample mean}}
The proof of the unbiasedness of the sample mean can be found in \cite[Thm. 2.1.1.]{Singh2003} and \cite[Thm. 2.2.1.]{Singh2003} for sampling with and without replacement, respectively.
Similarly, the sample mean variance given in~\eqref{eq:app_sample_mean_var}, obtained for either with and without replacement, is respectively proved in \cite[Thm. 2.1.2.]{Singh2003} and \cite[Thm. 2.2.3.]{Singh2003}.

\subsubsection{Proof of Lemma~\ref{lemma: reconstruction error}}
According to \eqref{eqn:SDQ}-\eqref{eqn:DiffDec}, it holds that
\begin{align*}    
    \ve^{\rm MEDU}_{u,t,m}=
    \begin{cases}
     \ve_{u,t,m}^{\rm SDQ} & m \in \bar{\cM}_t^u,\\
     \ve^{\rm thresh}_{u,t,m} & \text{otherwise;}
    \end{cases}    
\end{align*}
where $\ve_{u,t,m}^{\rm SDQ}$ is defined as in~\eqref{eq:sdq_error}, and $\ve^{\rm thresh}_{u,t,m}$ is given in~\eqref{itm:diff_thresh_noise}.
That is, the reconstruction error is either associated with \ac{sdq} or differential thresholding, and therefore, its expected value is anyhow zero from \ref{itm:zero_overeloading} and Theorem~\ref{thm:SDQ} or \ref{itm:diff_thresh_noise}, respectively. 

The  same arguments  hold for formulating the variance, i.e.,
\begin{align*}        
    \E\left[\left\|\ve^{\rm MEDU}_{u,t,m}\right\|^2\right] =
    \begin{cases}
     \E\left[\left\|\ve_{u,t,m}^{\rm SDQ}\right\|^2\right] = \sigma^2_{\rm SDQ} & m \in \bar{\cM}_t^u,\\
     \E\left[\left\|\ve^{\rm thresh}_{u,t,m}\right\|^2\right]\leq \delta_t^2 + \sigma^2_{\rm SDQ} & \text{otherwise;}
    \end{cases}  \\
     \leq\delta_t^2 + \sigma^2_{\rm SDQ}.   
\end{align*}
To justify the first inequality, note that:
    \begin{align*}
       & \E\Big[\big\|\ve^{\rm thresh}_{u,t,m}\big\|^2\Big] \overset{\eqref{itm:diff_thresh_noise}}{=} 
        \E\Big[\big\|\vg_{t,m}^u \!- \!\hat\vg_{{\tau_u(t)},m}^u \!+ \!\vg_{{\tau_u(t)},m}^u \!- \!\vg_{{\tau_u(t)},m}^u\big\|^2\Big] \\ 
        &\leq
        \E\Big[\big\|\vg_{t,m}^u \!- \!\vg_{{\tau_u(t)},m}^u\big\|^2\Big] \!+\! 
        \E\Big[\big\|\vg_{{\tau_u(t)},m}^u \!- \!\hat\vg_{{\tau_u(t)},m}^u\big\|^2\Big]\\ 
        &+2\E\left[\left\langle \vg_{t,m}^u \!- \!\hat\vg_{{\tau_u(t)},m}^u, \vg_{{\tau_u(t)},m}^u \!- \!\vg_{{\tau_u(t)},m}^u \right\rangle\right]        
        \overset{(a)}{\leq}\delta_t^2\! +\! \sigma^2_{\rm SDQ};
    \end{align*}
    where $(a)$ follows by \eqref{eqn:DiffThresh} and using the \ac{sdq} moments in~\eqref{eq:sdq_moments}.

Finally, the covariance obeys
\begin{align*}    
    \E\left[{\ve^{\rm MEDU}_{u,t,m}}^T
    \ve^{\rm MEDU}_{u',t',m}\right] =
    \begin{cases}
    \E\left[{\ve^{\rm SDQ}_{u,t,m}}^T
    \ve^{\rm SDQ}_{u',t',m}\right]=0,\\
    \E\left[{\ve^{\rm SDQ}_{u,t,m}}^T
    \ve^{\rm thresh}_{u',t',m}\right]=0,\\
    \E\left[{\ve^{\rm thresh}_{u,t,m}}^T
    \ve^{\rm thresh}_{u',t',m}\right]=0,\\
    \E\left[{\ve^{\rm thresh}_{u,t,m}}^T
    \ve^{\rm thresh}_{u,t',m}\right]\leq C/|t-t'|^\zeta.    
    \end{cases}   
\end{align*}
The above respectively follows as the \ac{sdq} errors are i.i.d zero-mean by Theorem~\ref{thm:SDQ} and the dither is independent of the source of randomness modeling the differential thresholding; the final bound stesm from the decaying-correlation errors by \ref{itm:diff_thresh_noise}.

\bibliographystyle{IEEEtran}
\bibliography{IEEEabrv,bib}

\end{document}

%% file: Preamble.tex
\usepackage{amsmath,amsfonts,bm}
\usepackage{graphicx}
\usepackage{acronym, enumitem}
\usepackage[linesnumbered,ruled,vlined]{algorithm2e}
\usepackage[noend]{algpseudocode}
\usepackage{amsthm}
\usepackage{tabularx}
\usepackage{pifont}
\usepackage{multirow}
\usepackage{multicol}
\usepackage{comment}
\usepackage{booktabs, makecell} 
\usepackage{flafter}
\usepackage{dsfont}

\usepackage[caption=false,font=normalsize,labelfont=sf,textfont=sf]{subfig}
\usepackage{textcomp}
\usepackage{stfloats}
\usepackage{verbatim}
\usepackage{cite}
\usepackage{hyperref}
\usepackage{array} 
\usepackage{url}
\usepackage{adjustbox}

\makeatletter
\newcommand{\leqnomode}{\tagsleft@true\let\veqno\@@leqno}%
\newcommand{\reqnomode}{\tagsleft@false\let\veqno\@@eqno}%
\newcommand*{\compress}{\@minipagetrue}
\makeatother

\acrodef{fl}[FL]{federated learning}
\acrodef{fu}[FU]{federated unlearning}
\acrodef{fa}[FedAvg]{federated averaging} 
\acrodef{mse}[MSE]{mean-squared error}
\acrodef{dq}[DQ]{dithered quantization}
\acrodef{sdq}[SDQ]{subtractive \ac{dq}}
\acrodef{nsdq}[NSDQ]{non-subtractive dithered quantization}
\acrodef{rtbf}[RTBF]{right to be forgotten}
\acrodef{mu}[MU]{machine unlearning}
\acrodef{sgd}[SGD]{stochastic gradient descent}
\acrodef{snr}[SNR]{signal-to-noise ratio}
\acrodef{rp}[RP]{random projection}
\acrodef{cnn}[CNN]{convolutional neural network}
\acrodef{medu}[MEDU]{memory-efficient distributed unlearning}

\newcommand{\cD}{\mathcal{D}}
\newcommand{\cU}{\mathcal{U}}
\newcommand{\cV}{\mathcal{V}}
\newcommand{\cL}{\mathcal{L}}
\newcommand{\cM}{\mathcal{M}}
\newcommand{\cP}{\mathcal{P}}

\newcommand{\cI}{\mathcal{I}}

\newcommand{\cX}{\mathcal{X}}

\def\1{\bm{1}}

\def\vc{{\bm{c}}}
\def\vd{{\bm{d}}}
\def\ve{{\bm{e}}}

\def\vg{{\bm{g}}}
\def\vh{{\bm{h}}}

\def\vl{{\bm{l}}}

\def\vw{{\bm{w}}}
\def\vx{{\bm{x}}}

\def\vz{{\bm{z}}}

\def\mG{{\bm{G}}}

\def\mI{{\bm{I}}}

\DeclareMathAlphabet{\mathsfit}{\encodingdefault}{\sfdefault}{m}{sl}
\SetMathAlphabet{\mathsfit}{bold}{\encodingdefault}{\sfdefault}{bx}{n}

\def\sR{{\mathbb{R}}}

\def\sZ{{\mathbb{Z}}}

\newcommand{\UnlUser}{\tilde{u}}

\newcommand{\E}{\mathbb{E}}

\newcommand{\Var}{\mathrm{Var}}

\newcommand{\Cov}{\mathrm{Cov}}

\DeclareMathOperator*{\argmin}{arg\,min}

\newtheorem{definition}{Definition}[section]
\newtheorem{observation}{Observation}[section]

\newtheorem{proposition}{Proposition}[section]
\newtheorem{lemma}{Lemma}[section]
\newtheorem{theorem}{Theorem}[section]
\newtheorem{corollary}{Corollary}[section]

%% file: Tikz_figs/Ablation_MNIST.tex
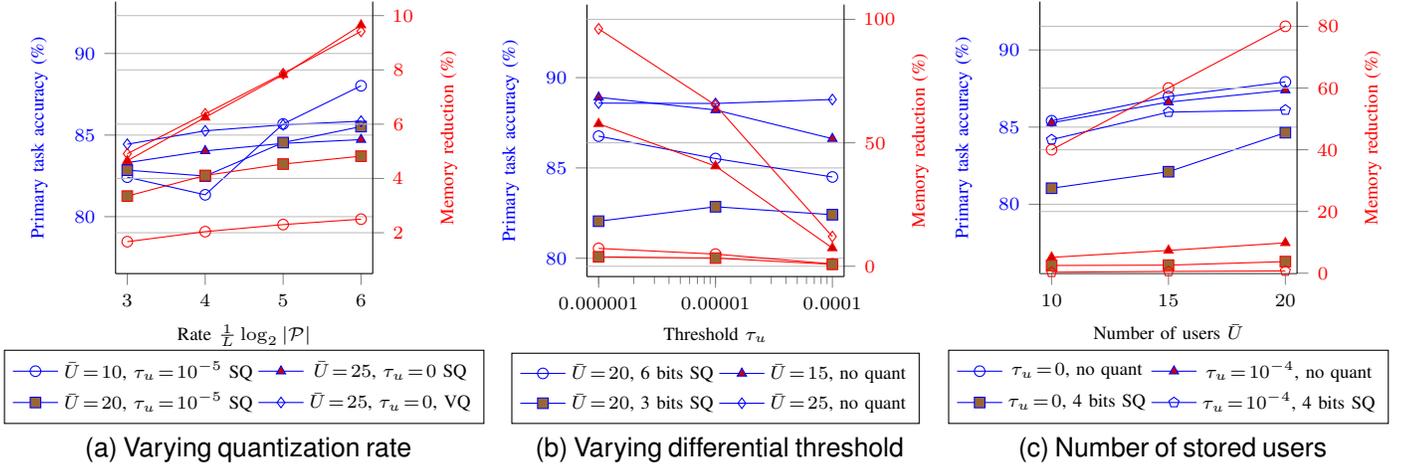
\begin{figure*}[t]
\centering
\subfloat[Varying quantization rate]{%
\begin{tikzpicture}
\begin{axis}[
  height=5.2cm,
  width=5cm,
  ylabel style={yshift=-0.9em, font=\scriptsize, color=blue},
  ylabel={Primary task accuracy (\%)},
  ymajorgrids=true,
  axis y line*=left,
  axis x line*=bottom,
  tick align=outside,
  tick label style={font=\scriptsize},
  yticklabel style={text width=2em,align=right},
  every y tick label/.append style={blue},
  ymin=77.27, ymax=92.42,
  enlargelimits=0.05,
  legend style={font=\scriptsize, at={(0.5,-0.28)},anchor=north,legend columns=2},
  xlabel={\scriptsize Rate $\frac{1}{L}\log_2|\mathcal{P}|$}
]
\addplot+[mark=o, solid, color=blue] coordinates {
(3, 82.42) (4, 81.34) (5, 85.68) (6, 88.02) };
\addlegendentry{$\bar{U}\!=\!10$, $\tau_u\!=\!10^{-5}$ SQ}
\addplot+[mark=triangle*, color=blue] coordinates {
(3, 83.31) (4, 84.03) (5, 84.49) (6, 84.72) };
\addlegendentry{$\bar{U}\!=\!25$, $\tau_u\!=\!0$ SQ}
\addplot+[mark=square*,  color=blue] coordinates {
(3, 82.85) (4, 82.49) (5, 84.54) (6, 85.52) };
\addlegendentry{$\bar{U}\!=\!20$, $\tau_u\!=\!10^{-5}$ SQ}
\addplot+[mark=diamond, color=blue] coordinates {
(3, 84.44) (4, 85.26) (5, 85.63) (6, 85.85) };
\addlegendentry{$\bar{U}\!=\!25$, $\tau_u\!=\!0$, VQ}
\end{axis}
\begin{axis}[
  height=5.2cm,
  width=5cm,
  axis y line*=right,
  axis x line=none,
  ylabel style={yshift=0.9em, font=\scriptsize, color=red},
  ylabel={Memory reduction (\%)},
  ymajorgrids=true,
  tick align=outside,
  tick label style={font=\scriptsize},
  yticklabel style={text width=2em,align=left},
  every y tick label/.append style={red},
  ymin=0.95, ymax=10.06,
  enlargelimits=0.05
]
\addplot+[mark=o, solid,   color=red] coordinates {
(3, 1.67) (4, 2.04) (5, 2.3) (6, 2.5) };
\addplot+[mark=triangle*,   color=red] coordinates {
(3, 4.65) (4, 6.25) (5, 7.82) (6, 9.65) };
\addplot+[mark=square*,   color=red] coordinates {
(3, 3.35) (4, 4.11) (5, 4.53) (6, 4.82) };
\addplot+[mark=diamond,    color=red] coordinates {
(3, 4.91) (4, 6.38) (5, 7.86) (6, 9.42) };
\end{axis}
\end{tikzpicture}
\label{fig:resultsA}
}
\subfloat[Varying differential threshold]{%
\begin{tikzpicture}
\begin{axis}[
  height=5.2cm,
  width=5cm,
  ylabel style={yshift=-0.9em, font=\scriptsize, color=blue},
  ylabel={Primary task accuracy (\%)},
  ymajorgrids=true,
  axis y line*=left,
  axis x line*=bottom,
  tick align=outside,
  tick label style={font=\scriptsize},
  yticklabel style={text width=2em,align=right},
  every y tick label/.append style={blue},
  ymin=79.67, ymax=93.36,
  enlargelimits=0.05,
  legend style={font=\scriptsize, at={(0.5,-0.28)},anchor=north,legend columns=2},
  xlabel={\scriptsize Threshold $\tau_u$},
  xmode=log,
  xtick={1e-6,1e-5,1e-4},
  log ticks with fixed point
]
\addplot+[mark=o, solid, color=blue] coordinates {
(1e-06, 86.76) (1e-05, 85.52) (0.0001, 84.5) };
\addlegendentry{$\bar{U}\!=\!20$, 6 bits SQ}
\addplot+[mark=triangle*, color=blue] coordinates {
(1e-06, 88.91) (1e-05, 88.21) (0.0001, 86.62) };
\addlegendentry{$\bar{U}\!=\!15$, no quant}
\addplot+[mark=square*,  color=blue] coordinates {
(1e-06, 82.05) (1e-05, 82.85) (0.0001, 82.41) };
\addlegendentry{$\bar{U}\!=\!20$, 3 bits SQ}
\addplot+[mark=diamond, color=blue] coordinates {
(1e-06, 88.59) (1e-05, 88.57) (0.0001, 88.79) };
\addlegendentry{$\bar{U}\!=\!25$, no quant}
\end{axis}
\begin{axis}[
  height=5.2cm,
  width=5cm,
  axis y line*=right,
  axis x line=none,
  ylabel style={yshift=0.9em, font=\scriptsize, color=red},
  ylabel={Memory reduction (\%)},
  ymajorgrids=true,
  tick align=outside,
  tick label style={font=\scriptsize},
  yticklabel style={text width=2em,align=left},
  every y tick label/.append style={red},
  ymin=0.83, ymax=100.97,
  enlargelimits=0.05,
  xmode=log,
  xtick=\empty
]
\addplot+[mark=o, solid,   color=red] coordinates {
(1e-06, 7.23) (1e-05, 4.92) (0.0001, 0.94) };
\addplot+[mark=triangle*,   color=red] coordinates {
(1e-06, 57.7) (1e-05, 40.56) (0.0001, 7.38) };
\addplot+[mark=square*,  color=red] coordinates {
(1e-06, 3.82) (1e-05, 3.3) (0.0001, 0.77) };
\addplot+[mark=diamond, color=red] coordinates {
(1e-06, 96.16) (1e-05, 65.23) (0.0001, 12.02) };
\end{axis}
\end{tikzpicture}
\label{fig:resultsB}
}
\subfloat[Number of stored users]{%
\begin{tikzpicture}
\begin{axis}[
  height=5.2cm,
  width=5cm,
  ylabel style={yshift=-0.9em, font=\scriptsize, color=blue},
  ylabel={Primary task accuracy (\%)},
  ymajorgrids=true,
  axis y line*=left,
  axis x line*=bottom,
  tick align=outside,
  tick label style={font=\scriptsize},
  yticklabel style={text width=2em,align=right},
  every y tick label/.append style={blue},
  ymin=76.31, ymax=92.33,
  enlargelimits=0.05,
  legend style={font=\scriptsize, at={(0.5,-0.28)},anchor=north,legend columns=2},
  xlabel={\scriptsize Number of users $\bar{U}$}
]
\addplot+[mark=o, solid,   color=blue] coordinates {
(10, 85.41) (15, 86.98) (20, 87.93) };
\addlegendentry{$\tau_u\!=\!0$, no quant}
\addplot+[mark=triangle*,   color=blue] coordinates {
(10, 85.28) (15, 86.62) (20, 87.4) };
\addlegendentry{$\tau_u\!=\!10^{-4}$, no quant}
\addplot+[mark=square*,  color=blue] coordinates {
(10, 81.04) (15, 82.11) (20, 84.64) };
\addlegendentry{$\tau_u\!=\!0$, 4 bits SQ}
\addplot+[mark=pentagon, color=blue] coordinates {
(10, 84.19) (15, 85.97) (20, 86.11) };
\addlegendentry{$\tau_u\!=\!10^{-4}$, 4 bits SQ}
\end{axis}
\begin{axis}[
  height=5.2cm,
  width=5cm,
  axis y line*=right,
  axis x line=none,
  ylabel style={yshift=0.9em, font=\scriptsize, color=red},
  ylabel={Memory reduction (\%)},
  ymajorgrids=true,
  tick align=outside,
  tick label style={font=\scriptsize},
  yticklabel style={text width=2em,align=left},
  every y tick label/.append style={red},
  ymin=3.88, ymax=84.00,
  enlargelimits=0.05
]
\addplot+[mark=o, solid,   color=red] coordinates {
(10, 40.0) (15, 60.0) (20, 80.0) };
\addplot+[mark=triangle*,   color=red] coordinates {
(10, 5.13) (15, 7.38) (20, 9.82) };
\addplot+[mark=square*,  color=red] coordinates {
(10, 2.5) (15, 2.61) (20, 3.75) };
\addplot+[mark=pentagon, color=red] coordinates {
(10, 0.32) (15, 0.58) (20, 0.72) };
\end{axis}
\end{tikzpicture}
\label{fig:resultsC}
}
\caption{Main accuracy and memory reduction on MNIST across varying parameters. Blue: accuracy (left); Red: memory (right).}
\label{fig:results}
\end{figure*}

%% file: Tikz_figs/Tradeoff_MNIST.tex
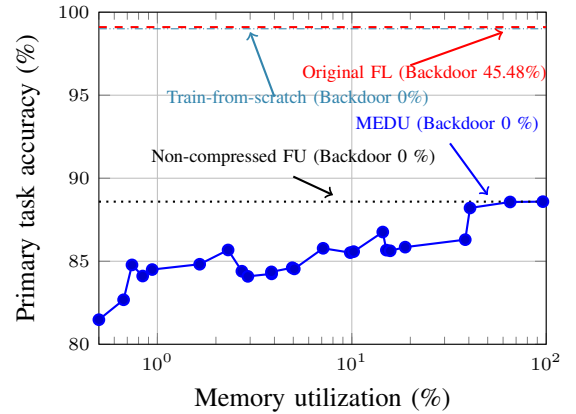
\begin{figure}[t]
\centering
\begin{tikzpicture}
\begin{axis}[
    width=0.85\linewidth,
    height=6cm,
    xlabel={Memory utilization (\%)},
    ylabel={Primary task accuracy (\%)},
    xmode=log,
    log basis x=10,
    xmin=0.5, xmax=100,
    ymin=80, ymax=100,
    grid=major,
    tick label style={font=\scriptsize},
]
\addplot+[color=blue, mark=*, thick] coordinates {
(0.50, 81.48)
(0.67, 82.68)
(0.74, 84.78)
(0.84, 84.11)
(0.94, 84.50)
(1.65, 84.82)
(2.31, 85.68)
(2.72, 84.40)
(2.92, 84.09)
(3.87, 84.24)
(3.86, 84.36)
(4.94, 84.61)
(5.06, 84.54)
(7.12, 85.78)
(9.82, 85.52)
(10.22, 85.58)
(14.43, 86.76)
(15.06, 85.67)
(15.76, 85.63)
(18.82, 85.85)
(38.34, 86.30)
(40.56, 88.21)
(65.23, 88.57)
(96.16, 88.59)
};

\addplot+[domain=0.5:100, samples=2, dashed, thick, color=red,mark=none, forget plot] {99.1};

\addplot+[domain=0.5:100, samples=2, densely dash dot dot, color=green!40!blue!80,mark=none, forget plot] {99};

\addplot+[domain=0.5:100, samples=2, dotted, thick, color=black, mark=none, forget plot] {88.59};

\node[anchor=west, font=\scriptsize, color=red] at (axis cs:5,96.3) {Original FL (Backdoor 45.48\%)};
\draw[->, thick, color=red] (axis cs:20,96.8) -- (axis cs:60,98.8);

\node[anchor=west, font=\scriptsize, color= green!40!blue!80] at (axis cs:1,94.8) {Train-from-scratch (Backdoor 0\%)};
\draw[->, thick, color=green!40!blue!80] (axis cs:4,94.9) -- (axis cs:3,98.8);

\node[anchor=south, font=\scriptsize, color=blue] at (axis cs:31,92.2) {MEDU (Backdoor 0 \%)};
\draw[->, thick, color=blue] (axis cs:32,92.1) -- (axis cs:50,88.99);

\node[anchor=south, font=\scriptsize, color=black] at (axis cs:5,90.2) {Non-compressed FU (Backdoor 0 \%)};
\draw[->, thick, color=black] (axis cs:5,90.1) -- (axis cs:8,88.99);

\end{axis}
\end{tikzpicture}
\caption{Memory-accuracy tradeoff for MEDU, MNIST.}
\label{fig:mnist_tradeoff}
\end{figure}

%% file: Tikz_figs/Tradeoff_CIFAR.tex
\begin{figure}[t]
\centering
\begin{tikzpicture}
\begin{axis}[
    width=0.85\linewidth,
    height=6cm,
    xlabel={Memory utilization (\%)},
    ylabel={Primary task accuracy (\%)},
    xmode=log,
    log basis x=10,
    xmin=1, xmax=100,
    ymin=30, ymax=72,
    grid=major,
    tick label style={font=\scriptsize},
]
\addplot+[color=blue, mark=*, thick] coordinates {
(1, 43.88)
(1.89, 44.46)
(2.6, 43.76)
(3.2, 43.24)
(3.71, 44.49)
(3.86, 43.94)
(4.34, 45.12)
(4.88, 45.13)
(5, 46.73)
(5.96, 46.82)
(6.17, 46.58)
(6.47, 47.96)
(7.22, 47.84)
(8.96, 47.26)
(9.24, 47.87)
(15.62, 47.82)
(18.75, 47.83)
(59.84, 45.22)
(75.83, 45.63)
(80, 45.85) 
(97.37, 45.63)
};

\addplot+[domain=1:100, samples=2, dashed, thick, color=red,mark=none, forget plot] {70.1};

\addplot+[domain=1:100, samples=2, densely dash dot dot, color=green!40!blue!80,mark=none, forget plot] {69.3};

\addplot+[domain=1:100, samples=2, dotted, thick, color=black, mark=none, forget plot] {45.63};

\node[anchor=west, font=\scriptsize, color=red] at (axis cs:8,65.3) {Original FL (Backdoor 13\%)};
\draw[->, thick, color=red] (axis cs:35,66.3) -- (axis cs:60,69.8);

\node[anchor=west, font=\scriptsize, color=green!40!blue!80] at (axis cs:1,61.0) {Train-from-scratch (Backdoor$\leq$1.3 \%)};
\draw[->, thick, color=green!40!blue!80] (axis cs:6,61.9) -- (axis cs:5,68.8);

\node[anchor=south, font=\scriptsize, color=blue] at (axis cs:21,54.2) {MEDU (Backdoor$\leq$1.3 \%)};
\draw[->, thick, color=blue] (axis cs:42,54.1) -- (axis cs:50,46.99);

\node[anchor=south, font=\scriptsize, color=black] at (axis cs:14,36.2) {Non-compressed FU (Backdoor$\leq$ 1.3\%)};
\draw[->, thick, color=black] (axis cs:14,40.1) -- (axis cs:20,45.99);

\end{axis}
\end{tikzpicture}
\caption{Memory-accuracy tradeoff for MEDU, CIFAR.}
\label{fig:cifar_tradeoff}
\end{figure}
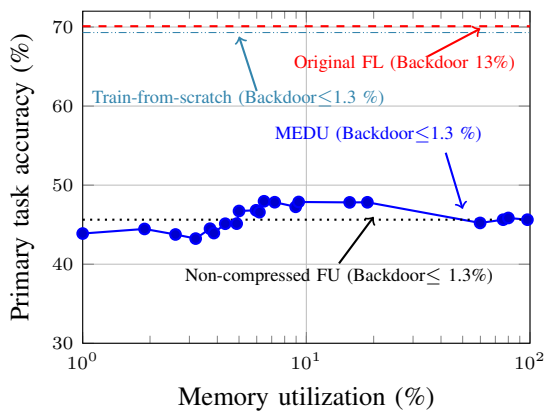